\documentclass[showpacs,preprintnumbers,amsmath,amssymb]{revtex4}
\usepackage[dvips]{graphicx}

\newcommand{\be}{\begin{equation}}
\newcommand{\ee}{\end{equation}}
\newcommand{\ba}{\begin{eqnarray}}
\newcommand{\ea}{\end{eqnarray}}

\newcommand{\Tr}[1]{ {\rm{Tr}} \! \left[ #1 \right] }
 
\newcommand{\difrac}[2]{{\displaystyle\frac{#1}{#2}}}
\def \Kbar{\overline{K}}

\begin{document}
\title{Scalar $\Lambda N$ and $\Lambda \Lambda$ interaction
       in a chiral unitary approach}

\author{K.~Sasaki}
\author{E.~Oset}
\author{M.~J.~Vicente Vacas}
 \affiliation{Departamento de F\'{\i}sica Te\'orica and IFIC,
Centro Mixto Universidad de Valencia-CSIC.}
 
\date{\today}

\begin{abstract}
We study the central part of $\Lambda N$ and $\Lambda \Lambda$ 
 potential by considering the correlated and uncorrelated two-meson
 exchange besides the $\omega$ exchange contribution.
The correlated two-meson is evaluated in a chiral unitary approach.
We find that a short range repulsion is generated by the correlated 
 two-meson potential which also produces an attraction in the
 intermediate distance region.
The uncorrelated two-meson exchange produces a sizeable attraction
 in all cases which is counterbalanced by $\omega$ exchange
 contribution.
\end{abstract}

\maketitle

\section{Introduction}

 The scalar isoscalar potential plays an important role in the nucleon nucleon 
 interaction providing an intermediate range attraction in all channels which 
 is demanded by the data. In models of the NN interaction using one boson
 exchange (OBE) this part of the interaction was accounted for by allowing the
 exchange of a "$\sigma$" particle in as early papers as \cite{hoshizaki}. The
 exchange of a scalar particle has been a constant in other OBE models 
 \cite{machleidt}.  In some models a broad $\varepsilon(760)$ was advocated 
 \cite{schwinger,binstock} which has been used later on in more recent models 
 \cite{nagels,rijken}. The particle data group \cite{PDG} refers to the 
 lightest scalar meson as $f_0(600)$ or $\sigma$. The nature of this particle,
 and even its mere existence, has been a source of controversy \cite{tornqvist},
 but in recent years a strong convergence towards the idea that the $\sigma$ is
 not a genuine meson state made up from a constituent $q \bar{q}$ pair has 
 been witnessed \cite{gasserulf,meissnersig}. This idea has obtained further
 support from unitarized chiral perturbation theory where the $\sigma$
 appears as a dynamically generated resonance from the $\pi \pi$ interaction
 \cite{dobado,dobadoram,ollernpa,oop,nsd,kaiserpi,markushin,nievesruiz,Guo:2005wp}. Other
 models that start from a seed of $q \bar{q}$, but couple this state to meson meson
 components in a unitary approach, converge to the same idea by showing that the
 meson cloud is essential in building up the low lying scalar resonances 
 \cite{tornqvistsig,vanBeveren,vanbdos}.
 
     The picture of the $\sigma$ as a dynamically generated resonance called
for a new interpretation of the $\sigma$ exchange in the NN interaction and
this work was performed in \cite{osettoki}. In this work the traditional 
$\sigma$ exchange was substituted by the exchange of two interacting mesons
within the chiral unitary framework of  \cite{ollernpa},  and an intermediate
attraction was found together with a repulsion at short distances, which makes the
picture qualitatively different from the ordinary, always attractive, 
$\sigma$ exchange.  The exchange of two interacting pions, although
nonperturbative,  was considered in \cite{gersten} and shown to reproduce well
the NN peripheral partial waves with $L> 2$. A recent work studying the isoscalar contact $NN$ interactions retakes the
unitarization of the $\pi \pi$ amplitude in the two pion exchange using the
Omnes representation \cite{donoghue}.

  The work of \cite{osettoki} was complemented in \cite{palomar}, where in
addition to the interacting two pion exchange, the contribution of the 
uncorrelated two pion exchange and the
repulsive contribution of the $\omega$ exchange were considered, leading
altogether to a good reproduction of the empirical scalar isoscalar interaction
of \cite{wiringa,wiringados}.

  The purpose of this work is to extend this to the strange sector evaluating
  the $\Lambda N$ scalar isoscalar interaction.
  
  Empirical evaluations of the YN scalar isoscalar interaction are done in
several works allowing the exchange of a scalar meson and making fits to data. 
As quoted above, the Nijmegen group makes use of the exchange of a heavy scalar
meson and there are different fits available in the literature
\cite{nagels,rijken,stoks,rijkensy}.  A recent work of the group shows an
interesting feature which is the improvement of the results by using a form
factor incorporating a zero \cite{rijkeny}, which leads to qualitative
features  of the scalar meson exchange similar to those obtained in
\cite{osettoki}. The exchange of various scalar mesons is also considered in
\cite{tominaga} as well as correlated two pion exchange, which however is
treated phenomenologically.  Another approach to the problem is the chiral quark
model in which the $\pi$ and a $\sigma$ are allowed to be exchanged between
constituent quarks \cite{straub,zhang}. In the same line, in the works of 
Refs. \cite{zhangdos,fujiwara} a SU(3) nonet of scalar mesons is exchanged
between the quarks.  

The closest work to our approach is
the theoretical work of \cite{haidenbauer}, following the line of the Juelich
model \cite{holinde,holindedos}, where the uncorrelated and
correlated two pion exchange are considered explicitly. The approach to the
correlated two pion (and two kaon) exchange is done rather differently by
evaluating theoretically the $B \bar{B} \to \pi \pi , K \bar{K}$ amplitudes and
using then unitarity and dispersion relations to relate  these amplitudes
to the correlated two meson exchange contribution to the $ BB \to BB$
interaction.  Our approach evaluates directly the correlated two pion
exchange by explicitly using the chiral unitary approach to deal with the pion
pion interaction and using appropriate  triangle diagrams to account for the
coupling of the two pions to the baryons.  The success of this approach
providing the scalar isoscalar $NN$ interaction provides solid grounds to extend
these ideas to the case of the $\Lambda N$ interaction, which we present in
this work.

\section{Correlated two-meson exchange between baryons}
\label{SEC:CorTM}
We follow closely Ref.~\cite{osettoki}
 and consider the correlated two-meson exchange between baryons.
In order to evaluate these diagrams 
 we use the lowest order chiral Lagrangeans
\begin{eqnarray}
{\cal{L}}_2
 &=& 
  \frac{1}{6f_\pi^2} \Tr{
     \Phi \partial^\mu \Phi \Phi \partial_\mu \Phi
   - \Phi \Phi \partial^\mu \Phi \partial_\mu \Phi
  }
+ \frac{1}{12f_\pi^2} \Tr{M \Phi^4 }
\\
 {\cal{L}}_{B} 
 &=& 
 \frac{D+F}{\sqrt{2} f_\pi} 
  \Tr{ \bar{B} \gamma_5 \gamma^\mu \partial_\mu \Phi B } 
 +
 \frac{D-F}{\sqrt{2} f_\pi}  
  \Tr{ \bar{B} \gamma_5 \gamma^\mu B \partial_\mu \Phi } 
\end{eqnarray}
where $\Phi$ and $B$ are the standard $SU(3)$ matrices
 for the octets of pseudoscalar mesons and baryons
 respectively
 \cite{Gasser:1984gg,Gasser:1984ux,Gasser:1984pr,
       Ecker:1994gg,Meissner:1993ah,Bernard:1995dp,
       Oller:2000ma}.
The mass matrix of the mesons octet is defined by 
 $M \equiv {\textrm{diag}}(m_\pi^2, m_\pi^2, 2m_K^2-m_\pi^2)$.
From there,
 one obtains the different pion-pion (or $K \Kbar$) lowest 
 order amplitudes, which can be found in \cite{ollernpa}
 and the $MBB$ vertices which for convenience we show 
 in the appendix.

The isoscalar amplitudes, which contain only $s$-wave,
 have been obtained by employing Lagrangean ${\cal{L}}_2$ 
 in \cite{oop}.
The lowest order, tree level, amplitudes of meson-meson
 scattering can be written as
\begin{eqnarray}
&&
 t^{(I=0,L)}_{\pi \pi \to \pi \pi}
  = 
  - \frac{1}{f_\pi^2} \left( s - \frac{m_\pi^2}{2} \right)
  + \frac{1}{3f_\pi^2} \sum_i ( p_i^2 - m_i^2 )
\label{EQ:4} \\
&&
 t^{(I=0,L)}_{K \Kbar \to K \Kbar}
  =
 - \frac{3s}{4f_\pi^2}
 + \frac{1}{4 f_\pi^2} \sum_i ( p_i^2 - m_i^2 )
\label{EQ:AK} \\
&&
 t^{(I=0,L)}_{\pi \pi \to K \Kbar}
  =
 - \frac{\sqrt{3} s}{4 f_\pi^2}
 + \frac{1}{4 \sqrt{3} f_\pi^2} \sum_i ( p_i^2 - m_i^2 )
\label{EQ:ApK}
\end{eqnarray}
where the $L$ on the superscript stands for the leading order
 amplitude of meson-meson scattering and we have employed the
 convenient unitary normalization and the isospin phase
 convention 
 ($| \pi^+ \rangle = -|1,1 \rangle$, 
  $| K^- \rangle = -|1/2,-1/2 \rangle$):
\begin{eqnarray}
&&
 | \pi \pi, (I=0) \rangle
  = - \frac{1}{\sqrt{6}}
  | \pi^0 \pi^0 + \pi^+ \pi^- + \pi^- \pi^+ \rangle
\\
&&
| K \Kbar (I=0) \rangle
  = - \frac{1}{\sqrt{2}}|K^0 \Kbar^0 + K^+ K^- \rangle.
\end{eqnarray}
In eqs.~(\ref{EQ:4},\ref{EQ:AK},\ref{EQ:ApK})
 we have separated the lowest order interaction into a part
 which provides the on-shell contribution
 and another term (the one with $(p_i^2 - m_i^2)$) 
 which contributes only for off-shell mesons.

\begin{figure}
 \begin{center}
\includegraphics[scale=0.85]{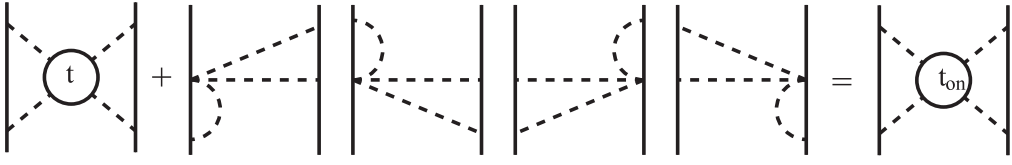}
 \end{center}
\caption{Set of diagrams that cancel the off shell part
         of the correlated two-meson exchange contribution.}
\label{Fig:Off-Shell-C}
\end{figure}
As shown in \cite{oop,osettoki,palomar},
 the off-shell part of the meson-meson amplitudes does not
 contribute to our calculation.
In fact,
 for the meson-meson loops,
 this contribution is absorbed into the physical mass and 
 the coupling.
As for the coupling to the baryons,
 there is a cancellation of the off-shell part of the meson-meson
 amplitude in eq.~(\ref{EQ:4},\ref{EQ:AK},\ref{EQ:ApK})
 with the diagrams of the type of Fig.~\ref{Fig:Off-Shell-C}.
This fact is valid not only for the $NN$ case but also for 
 $YN$ and $YY$ case.
Thus, hereafter, 
 we only consider the on-shell part of the meson-meson amplitude.

\begin{figure}
\begin{tabular}{cp{1em}cp{1em}c}
\includegraphics[scale=.22]{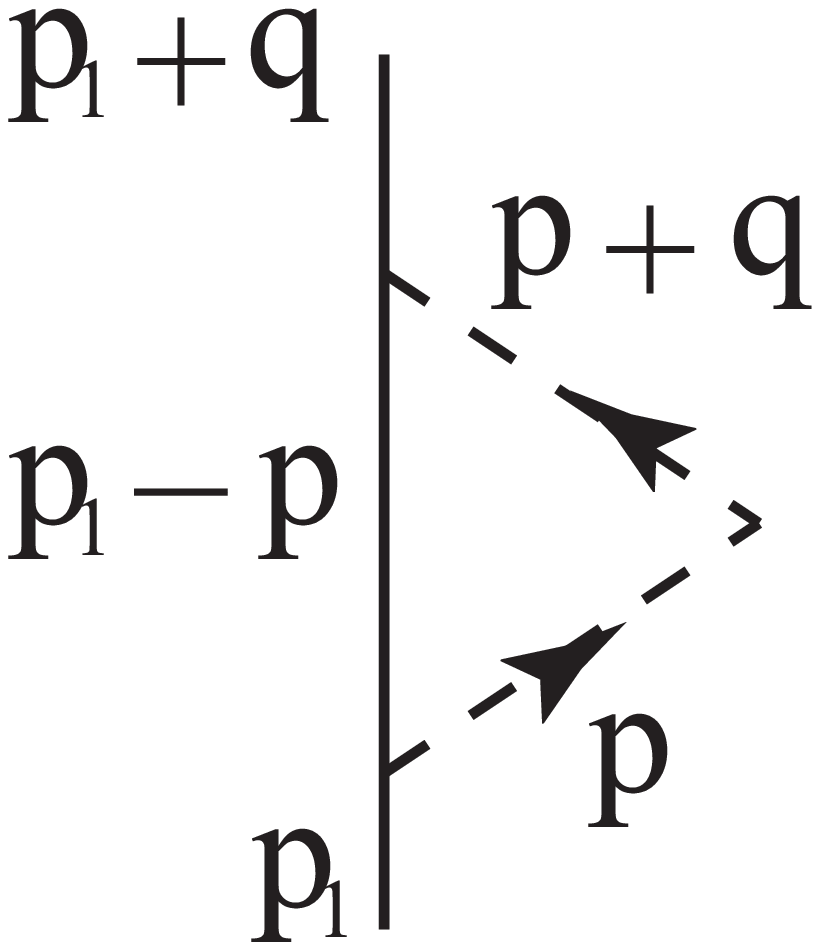}  & &
\includegraphics[scale=.22]{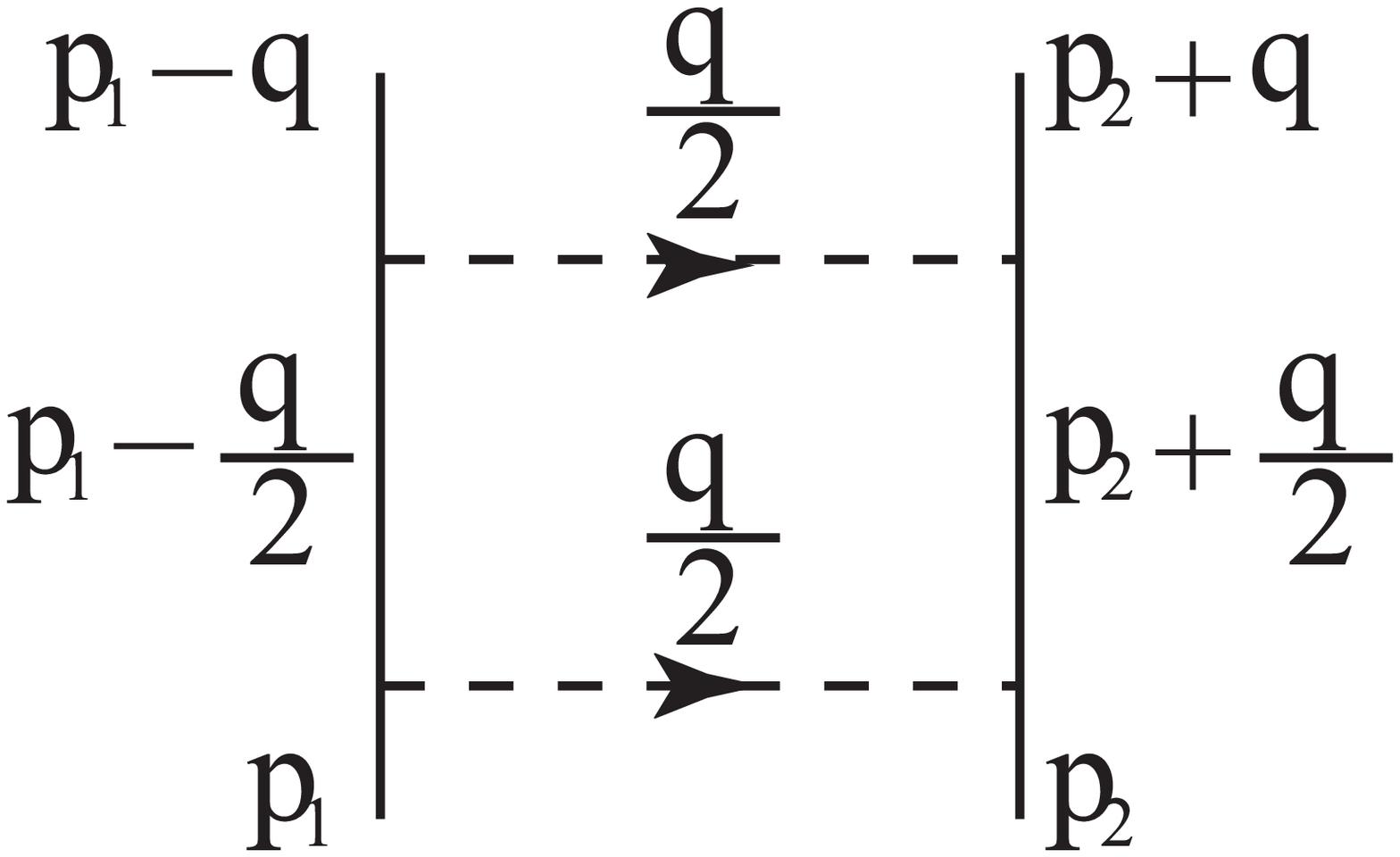} & &
\includegraphics[scale=.22]{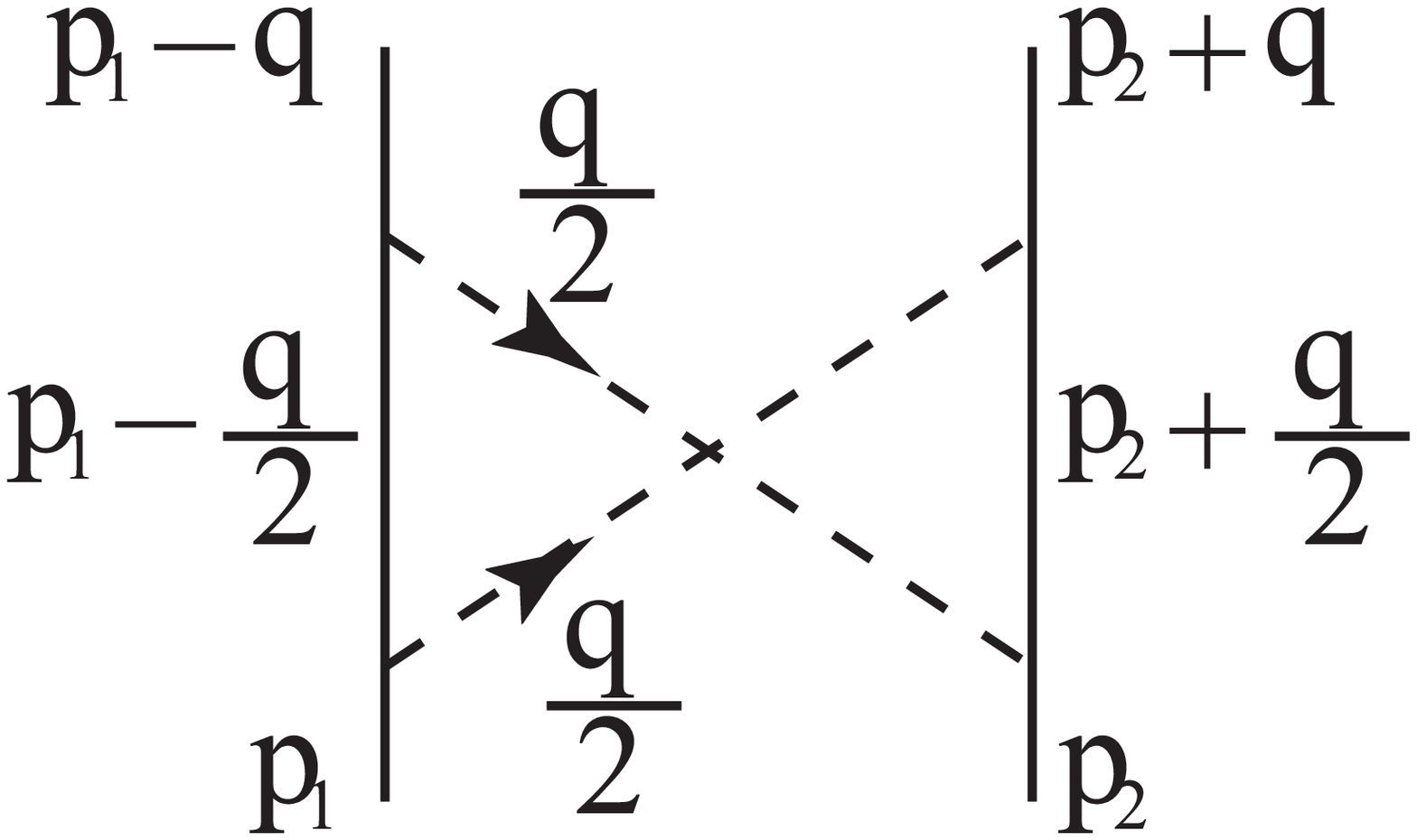} \\
(a) & & (b) & & (c)
\end{tabular}
\caption{}
\label{Fig:MomC}
\end{figure}
This on-shell treatment enables us to separate the on-shell
 meson-meson amplitude from the triangle loop integration
 that couples the mesons to the baryons.
Thus we can define the correlated two-meson potential as
\begin{eqnarray}
 V^{Cor}_{B_1 B_2}(q)
  = \sum_{ij}^{\pi \pi, K \Kbar}
    N_{ij}
    \Delta_{B_1}^i t^{(I=0,L)}_{i \to j} \Delta_{B_2}^j
\end{eqnarray}
where $\Delta$ indicates the triangle loop contribution of two-meson
 for baryon $B_k$ and $N_{ij}$ is a factor from the isospin 
 summation, concreately $N_{\pi \pi, \pi \pi}=6$, 
 $N_{\pi \pi, K \Kbar} = N_{K \Kbar, \pi \pi} = 2 \sqrt{3}$,
 $N_{K \Kbar, K \Kbar} = 2$.
For concreteness,
 the $\Delta$ function in the correlated two-pion potential for $NN$
 channel is given by
\begin{eqnarray}
 \Delta^{(\pi \pi)}_N
 & = &
    \left( \frac{D+F}{2 f_\pi} \right)^2
    V^{(\pi \pi)}_{N N} (q)
\end{eqnarray}
where $V^{(m_1 m_2)}_{B'B}(q)$ is the vertex function which is
 already evaluated in \cite{osettoki} and given in a generalized
 form as
\begin{eqnarray}
  V^{(m_1 m_2)}_{B'B}(q) 
  = 
   \int \frac{d^3p}{(2 \pi)^3}
   \frac{M_{B'}}{E_{B'}(\vec{p})}
    \frac{ ( \vec{p} + \vec{q} ) \cdot \vec{p} }
         { 2 \omega_1 \omega_2 ( \omega_1 + \omega_2 )} 
    \frac{  \omega_1 + \omega_2 + E_{B'}(\vec{p})-M_B }
         { ( \omega_1 + E_{B'}(\vec{p})-M_B )
           ( \omega_2 + E_{B'}(\vec{p})-M_B )}
\label{EQ:Triamp}
\end{eqnarray}
 with 
\begin{eqnarray}
 E_{B'}(\vec{p}) = \sqrt{\vec{p}^2+M_{B'}^2}; \hspace*{1em}
 \omega_1 = \sqrt{\mu^2_1 + \vec{p}^2}; \hspace*{1em}
 \omega_2 = \sqrt{\mu^2_2 + (\vec{p} + \vec{q})^2}.
\end{eqnarray}
This is calculated with the variables corresponding to
 Fig.~\ref{Fig:MomC}(a)
 and where, as in \cite{osettoki}, we have put the initial momentum
 at rest.
We introduce a static form factor in order to regularize 
 the triangle loop function.
The form factor employed in this calculation is
\begin{eqnarray}
 F(\vec{p}) F(\vec{p}+\vec{q})
 = 
 \frac{\Lambda^2}{\Lambda^2 + \vec{p}^2}
 \frac{\Lambda^2}{\Lambda^2 + (\vec{p}+\vec{q})^2}
\end{eqnarray}
where the cutoff is chosen as $\Lambda = 1.0~{\rm{GeV}}$.


\subsection{Lowest order contribution in the isoscalar exchange 
         in the $\Lambda N \to \Lambda N$ interaction}

%
\begin{figure}
\includegraphics[scale=.85]{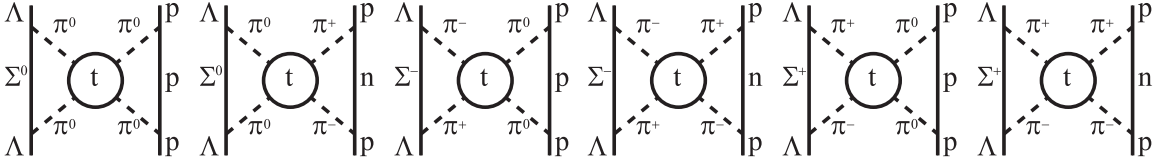}
\caption{Diagrams of scalar-isoscalar $\Lambda N$ processes
         involving $t_{\pi \pi \to \pi \pi}$}
\label{Fig:1}
\end{figure}
For this case
 the potential generated by the correlated two-pion diagrams
 shown in Fig.~\ref{Fig:1} is given by
\begin{eqnarray}
 V^{\pi \pi \to \pi \pi} 
  = 
 6
 \left[
  \left( \frac{D}{\sqrt{3} f_\pi} \right)^2
  V^{(\pi \pi)}_{\Sigma \Lambda} (q)
 \right]
 t^{(I=0,L)}_{\pi \pi \to \pi \pi}
 \left[
  \left( \frac{D+F}{2 f_\pi} \right)^2
  V^{(\pi \pi)}_{N N} (q)
 \right]
\label{EQ:pipioct}
\end{eqnarray}
where $V^{(\pi \pi)}_{\Sigma \Lambda} (q)$ is the vertex function
 defined in eq.~(\ref{EQ:Triamp}) with the same form factor.
Furthermore, as discribed in section~\ref{SEC:UNIT}, 
 we substitute $t^{(I=0,L)}_{\pi \pi \to \pi \pi}$
 by the full unitarized amplitude.
From now on
 we consider directly the full meson-meson amplitudes in all cases.

%
\begin{figure}
\begin{center}
 \includegraphics[scale=.85]{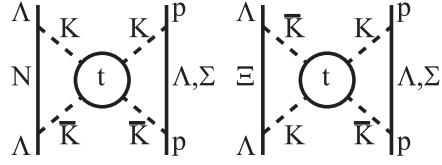}
\end{center}
\caption{Diagrams involving $t_{K \Kbar \to K \Kbar}$}
\label{Fig:3}
\end{figure}
%
We also include the exchange of $K \Kbar$ in the approach.
The diagrams to take into account are shown in Fig.~\ref{Fig:3}.
By using the couplings in the appendix,
 we evaluate the potential generated by the correlated $K \Kbar$ 
 contributions in a similar way as before obtaining
\begin{eqnarray}
 V^{K \Kbar \to K \Kbar} 
  & = &
 2
 \Delta^{K \Kbar}_\Lambda ~
   t^{(I=0,L)}_{K \Kbar \to K \Kbar} ~
 \Delta^{K \Kbar}_N 
\end{eqnarray}
with the triangle kaon-loop contribution given by
\begin{eqnarray}
 \Delta^{(K \Kbar)}_\Lambda
  & = &
   \left( \frac{D+3F}{2 \sqrt{3} f_\pi} \right)^2
   V^{(K \Kbar)}_{N \Lambda} (q)
 + \left( \frac{3F-D}{2 \sqrt{3} f_\pi} \right)^2
   V^{(K \Kbar)}_{\Xi \Lambda} (q)
\\
 \Delta^{(K \Kbar)}_N
  & = &
   \frac{3}{2}
   \left( \frac{D-F}{2 \sqrt{3} f_\pi} \right)^2
   V^{(K \Kbar)}_{\Sigma N} (q)
 + \frac{1}{2}
   \left( \frac{D+3F}{2 \sqrt{3} f_\pi} \right)^2
   V^{(K \Kbar)}_{\Lambda N} (q)
\end{eqnarray}
The factors in the brace of $\Delta^{(K \Kbar)}_N$ 
 come from the $I=0$ projection of the kaon-couplings.

%
\begin{figure}
\begin{center}
 \includegraphics[scale=.85]{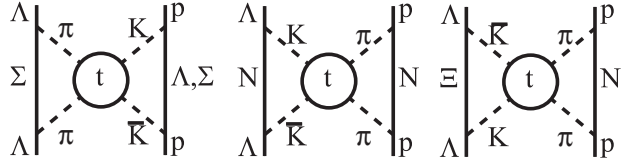}
\end{center}
\caption{Diagrams involving $t_{\pi \pi \to K \Kbar}$}
\label{Fig:4}
\end{figure}
Now one must consider the mixed terms with vertices with 
 $\pi$ or $K$ which involve the $\pi \pi \to K \Kbar$
 transition amplitude.
The diagrams to consider are shown in Fig.~\ref{Fig:4}.
The potential in this case is given by 
\begin{eqnarray}
 V^{\pi \pi \to K \Kbar} 
  & = &
 2 \sqrt{3}
 \Delta^{\pi \pi}_\Lambda ~
   t^{(I=0,L)}_{\pi \pi \to K \Kbar} ~
 \Delta^{K \Kbar}_N 
 +
 2 \sqrt{3}
 \Delta^{K \Kbar}_\Lambda ~
   t^{(I=0,L)}_{K \Kbar \to \pi \pi} ~
 \Delta^{\pi \pi}_N 
\end{eqnarray}
with the triangle meson-loop contribution shown before.

\subsection{Contribution of $\Delta$, $\Sigma^\ast$, $\Xi^\ast$
            intermediate states}
Next 
 we wish to include the contribution of the intermediate 
 $\Delta$, $\Sigma^\ast$, $\Xi^\ast$ states.
In the block of diagrams of Fig.~\ref{Fig:1}
 we can introduce $\Sigma^\ast$ in the left triangular 
 vertex, or $\Delta$ in the right triangular vertex, or both.

The coupling of the decuplet to the octet of mesons and baryons
 is given by 
\begin{eqnarray}
 {\cal{L}}^{Dec}
  &=& 
    \frac{ \sqrt{2} }{ f_\pi } {\cal{C}}
    \sum_{a,b,c,d,e}^{1 \sim 3}
     \epsilon^{abc}
    \left(
     ( \bar{T}_{ade} \Phi^d_b {B}^e_c ) 
      \vec{S} \cdot (- \vec{q})
   + ( \bar{B}^c_e \Phi^b_d T_{ade} )
      \vec{S}^\dagger \cdot \vec{q}
    \right)
\end{eqnarray}
for an outgoing meson with momentum $q$.
The ${\cal{C}}$ is determined from 
 the $\Delta N \pi$ coupling constant.
$T$ is the decuplet baryon field shown in the appendix.
This Lagrangean gives rise to couplings of the type
\begin{eqnarray}
 \gamma_{MBB'}~\frac{f_{\pi N \Delta}^\ast}{m_\pi}~ 
 \vec{S} \cdot \vec{q}
\end{eqnarray}
for outgoing mesons of momentum $\vec{q}$.
The $\gamma_{MBB'}$ coefficients can be found in the appendix.
We use $f_{\pi N \Delta}^\ast=2.12f_{\pi NN}$ to obtain 
 the correct width of the $\Delta(1232)$ resonance.

Altogether, the contribution involving
 the $\pi \pi \to \pi \pi$ amplitude is given by
\begin{eqnarray}
 V^{(\pi \pi \to \pi \pi)}
 & = &
 6~t^{(I=0,L)}_{\pi \pi \to \pi \pi}
 \left[ 
   \left( \frac{D}{\sqrt{3} f_\pi} \right)^2 
    V^{(\pi \pi)}_{\Sigma \Lambda} (q)
 + \frac{2}{3}
   \left( \frac{f_{\pi N \Delta}^\ast}{\sqrt{2} m_\pi} \right)^2 
    V^{(\pi \pi)}_{\Sigma^\ast N} (q)
 \right]
 \nonumber \\
 &   & \hspace*{3em} \times
 \left[ 
   \left( \frac{D+F}{2 f_\pi} \right)^2 
    V^{(\pi \pi)}_{N \Lambda} (q)
 + \frac{4}{9} \left( \frac{f_{\pi N \Delta}^\ast}{m_\pi} \right)^2 
    V^{(\pi \pi)}_{\Delta N} (q)
 \right]
\label{EQ:FullCpp}
\end{eqnarray}
where the factor two-thirds comes from the difference of spin
 and the extra two-thirds in front of $V^{(\pi \pi)}_{\Delta N}$
 from the change of isospin.
This equation shows how the triangle loop contribution is modified 
 by the excited baryon in the intermediate state.
Thus, the modified triangle loop contribution can be written as
\begin{eqnarray}
 \tilde{\Delta}^{(\pi \pi)}_{\Lambda}
 & = &
   \Delta^{(\pi \pi)}_{\Lambda}
 + 
   \frac{2}{3}
   \left( \frac{f_{\pi N \Delta}^\ast}{\sqrt{2} m_\pi} \right)^2 
    V^{(\pi \pi)}_{\Sigma^\ast N} (q)
\\
 \tilde{\Delta}^{(\pi \pi)}_N
 & = &
 \Delta^{(\pi \pi)}_N
 + 
    \frac{4}{9} \left( \frac{f_{\pi N \Delta}^\ast}{m_\pi} \right)^2 
    V^{(\pi \pi)}_{\Delta N} (q).
\end{eqnarray}

In the same way 
 we take into account the contributions of $\Sigma^\ast$,
 $\Xi^\ast$ in the $K \Kbar$ triangle contribution by means of 
 the couplings of the appendix, and the results are given by
\begin{eqnarray}
 \tilde{\Delta}^{(K \Kbar)}_\Lambda
  & = &
 \Delta^{(K \Kbar)}_\Lambda
 + 
  \frac{2}{3}
  \left( \frac{f_{\pi N \Delta}^\ast}{\sqrt{2} m_\pi} \right)^2 
   V^{(K)}_{\Xi^\ast} (q)
\\
 \tilde{\Delta}^{(K \Kbar)}_N
  & = &
 \Delta^{(K \Kbar)}_N
 + 
   \frac{2}{3}
   \left( \frac{f_{\pi N \Delta}^\ast}{\sqrt{6} m_\pi} \right)^2 
    V^{(K)}_{\Sigma^\ast} (q).
\end{eqnarray}

Therefore 
 the total leading-order potential generated by the correlated 
 two-meson contribution with the excited baryons in the intermediate 
 states is given by substitution of the $\Delta$ to 
 the $\tilde{\Delta}$ as
\begin{eqnarray}
 V^{Cor}_{\Lambda N} (q)
  & = &
 \sum_{i,j}^{ \pi \pi, K \Kbar}
 N_{ij}
 \tilde{\Delta}^{i}_\Lambda ~
   t^{(I=0,L)}_{i \to j} ~
 \tilde{\Delta}^{j}_N .
\label{EQ:Fullrep}
\end{eqnarray}

\subsection{Unitarization of the amplitudes}
\label{SEC:UNIT}
Here we follow Ref.~\cite{ollernpa} and iterate the meson-meson
 potential to infinite order by using a Lippmann-Schwinger type 
 equation in coupled $\pi \pi$ and $K \Kbar$ channels.
As it is shown in \cite{ollernpa,oop},
 the Lippmann-Schwinger equation can be reduced under the
 on-shell factorization to the algebraic relation
\begin{eqnarray}
 T = \left[ 1 - V G \right]^{-1} V
\end{eqnarray}
where $V \equiv t$ used in the former 
 eqs.(\ref{EQ:4},\ref{EQ:AK},\ref{EQ:ApK}), and
 $G$ is the meson-meson loop function.
The $G$ function is given by 
\begin{eqnarray}
 G(s) 
  = 
 \int_0^{q_{\rm{max}}} \frac{q^2 dq}{(2 \pi)^2}
 \frac{\omega_1 + \omega_2}
      {\omega_1 \omega_2 [s - (\omega_1 + \omega_2)^2 + i \epsilon]}
\end{eqnarray}
where $\omega_i = \sqrt{\vec{q}^2 + m_i^2}$.
It is regularized with the cutoff scheme
 which is different from the expression used in the previous paper 
 \cite{osettoki}.
One advantage of the usage of a cutoff is that it does not 
 produce undesirable poles as mentioned in \cite{osettoki}.
For the region of interest to us, one can use both cutoff or 
 dimensional regularization to evaluate $G$.
Some caveats about the use at unreasonably low negative values of $s$
 are discussed in \cite{osettoki}, which set restrictions on the
 results at very short distances.

We have also included the $\eta \eta$ channel but their effect is
 small and can be approximately reabsorbed in the $\pi \pi$,
 $K \Kbar$ channels by redefining $q_{\rm{max}}$ 
 \cite{kaiserpi}.
We obtain good results for the $\pi \pi$ phase shift up to
 $1.2~{\rm{GeV}}$ by using the $\pi \pi$, $K \Kbar$ channels
 and $q_{\rm{max}} = 1.0~{\rm{GeV}}$.


The final expression for the $\Lambda N$ scalar potential is 
 given by summing the expressions in section \ref{SEC:CorTM} 
 eq.~(\ref{EQ:Fullrep}) and substituting $t_i$ by the unitarized
 amplitude $T_i$.

\section{Uncorrelated two meson exchange}
\begin{figure}
\begin{center}
 \includegraphics[scale=.85]{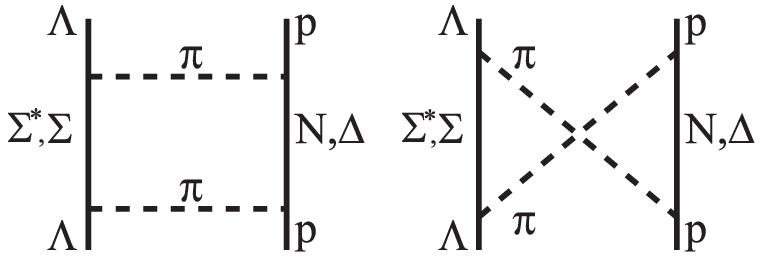} 
 \hspace*{0.5em}
 \includegraphics[scale=.85]{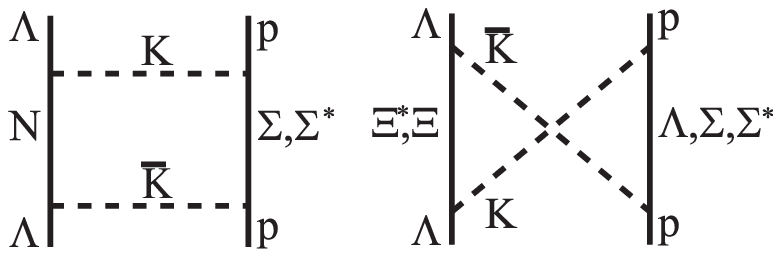}
\end{center}
\caption{Uncorrelated two-pion and two-kaon diagrams}
\label{Fig:6}
\end{figure}
We consider both the direct and crossed diagrams of the
 uncorrelated two-pion and two-kaon exchange contributions
 shown in Fig.~\ref{Fig:6}
We follow here the procedure of \cite{palomar} and use
 the variables of the diagrams as shown in
 Fig.~\ref{Fig:MomC}(b), (c).

The uncorrelated two-meson potential is given after performing 
 analytically the $p^0$ integration in terms of the integrals
\begin{eqnarray}
 V^{(i,M_1 M_2)}_{B_1 B_2}(q)
  = -
 \int \frac{d^3p}{(2 \pi)^3} \frac{M_1}{E_1} \frac{M_2}{E_2}
 \left( p^2 - \frac{q^2}{4} \right)^2
 R_i(\cdot)
\end{eqnarray}
where $M_1$ and $M_2$ are the baryon masses and 
 $i$ stands for the direct $(D)$ or crossed $(C)$ terms
 and
\begin{eqnarray}
 R_D(\cdot)
  & = &
  \frac{
   (\omega_1 + \omega_2)(( E'_1 + E'_2 )^2 + 2 \omega_1 \omega_2)
 + (\omega_1^2 + 3 \omega_1 \omega_2 + \omega_2^2
   + E'_1 E'_2 ) ( E'_1 + E'_2 )}
 { 2 \omega_1 \omega_2 ( \omega_1 + \omega_2 )
   ( E'_1 + E'_2 )
   ( \omega_1 + E'_1 )
   ( \omega_1 + E'_2 ) 
   ( \omega_2 + E'_1 )
   ( \omega_2 + E'_2 )}
\\
 R_C(\cdot)
  & = &
  \frac{
   (\omega_1 + \omega_2)( E'_1 + E'_2 )
 +  \omega_1^2 + \omega_1 \omega_2 + \omega_2^2
   + E'_1 E'_2 }
 { 2 \omega_1 \omega_2 ( \omega_1 + \omega_2 )
   ( \omega_1 + E'_1 )
   ( \omega_1 + E'_2 ) 
   ( \omega_2 + E'_1 )
   ( \omega_2 + E'_2 )}
\end{eqnarray}
with $E'_i = E_i - p_i^0$.
It is worth mentioning that, if we compare without coupling
 constants, the crossed contribution is much smaller than
 the box type contribution.
Furthermore, 
 both $V^D$ and $V^C$ are largely suppressed by the mass of the
 exchanged meson.
As a result of this rough estimation, 
 we expect that
 the uncorrelated two-kaon contribution is much smaller than
 the uncorrelated two-pion contribution.

By taking the coupling of the vertices in the different 
 diagrams into account 
 we get the contribution to the $\Lambda N$ potential 
 with considering the contribution of the
 $\Delta$, $\Sigma^\ast$, $\Xi^\ast$ in the intermediate
 states as
\begin{eqnarray}
v^{(D,\pi \pi)}_{\Lambda N}
&=&
 3
 \left( \frac{D}{\sqrt{3} f_\pi} \right)^2
 \left( \frac{D+F}{2 f_\pi} \right)^2
 V^{(D,\pi \pi)}_{\Sigma N} (q)
 + 
 2
 \left( \frac{D+F}{2 f_\pi} \right)^2
 \left( \frac{f_{\pi N \Delta}^\ast}{\sqrt{2} m_\pi} \right)^2
 V^{(D,\pi \pi)}_{\Sigma^\ast N} (q)
\nonumber \\
&&
 +
 \frac{4}{3}
 \left( \frac{D}{\sqrt{3} f_\pi} \right)^2
 \left( \frac{f_{\pi N \Delta}^\ast}{m_\pi} \right)^2
 V^{(D,\pi \pi)}_{\Sigma \Delta} (q)
 + 
 \frac{8}{9}
 \left( \frac{f_{\pi N \Delta}^\ast}{\sqrt{2} m_\pi} \right)^2
 \left( \frac{f_{\pi N \Delta}^\ast}{m_\pi} \right)^2
 V^{(D,\pi \pi)}_{\Sigma^\ast \Delta} (q)
\nonumber \\
v^{(C,\pi \pi)}_{\Lambda N}
&=&
 3
 \left( \frac{D}{\sqrt{3} f_\pi} \right)^2
 \left( \frac{D+F}{2 f_\pi} \right)^2
 V^{(C,\pi \pi)}_{\Sigma N} (q)
 +
 2
 \left( \frac{f_{\pi N \Delta}^\ast}{\sqrt{2} m_\pi} \right)^2
 \left( \frac{D+F}{2 f_\pi} \right)^2
 V^{(C,\pi \pi)}_{\Sigma^\ast N} (q)
\nonumber \\
&&
 +
 \frac{4}{3}
 \left( \frac{D}{\sqrt{3} f_\pi} \right)^2
 \left( \frac{f_{\pi N \Delta}^\ast}{m_\pi} \right)^2
 V^{(C,\pi \pi)}_{\Sigma \Delta} (q)
 +
 \frac{8}{9}
 \left( \frac{f_{\pi N \Delta}^\ast}{\sqrt{2} m_\pi} \right)^2
 \left( \frac{f_{\pi N \Delta}^\ast}{m_\pi} \right)^2
 V^{(C,\pi \pi)}_{\Sigma^\ast \Delta} (q)
\end{eqnarray}
where we write explicitly the left and right baryon in the 
 diagrams and the couple of mesons exchanged.

Similarly for the $K \Kbar$ diagrams we calculate the uncorrelated
 two-kaon potential as
\begin{eqnarray}
v^{(D,K \Kbar)}_{\Lambda N}
& = &
 3
 \left( \frac{D+3F}{2 \sqrt{3} f_\pi} \right)^2
 \left( \frac{D-F}{2 f_\pi} \right)^2
 V^{(D,K \Kbar)}_{N \Sigma} (q)
 + 
 \frac{1}{3}
 \left( \frac{D+3F}{2 \sqrt{3} f_\pi} \right)^2
 \left( \frac{f_{\pi N \Delta}^\ast}{m_\pi} \right)^2
 V^{(D,K \Kbar)}_{N \Sigma^\ast} (q)
\nonumber \\
v^{(C,K \Kbar)}_{\Lambda N}
& = &
 3
 \left( \frac{3F-D}{2 \sqrt{3} f_\pi} \right)^2
 \left( \frac{D-F}{2 f_\pi} \right)^2
 V^{(C,\Kbar K)}_{\Xi \Sigma} (q)
 +
 2
 \left( \frac{f_{\pi N \Delta}^\ast}{\sqrt{2} m_\pi} \right)^2
 \left( \frac{D-F}{2 f_\pi} \right)^2
 V^{(C,\Kbar K)}_{\Xi^\ast \Sigma} (q)
\nonumber \\
&&
 +
 2
 \left( \frac{3F-D}{2 \sqrt{3} f_\pi} \right)^2
 \left( \frac{f_{\pi N \Delta}^\ast}{\sqrt{6} m_\pi} \right)^2
 V^{(C,\Kbar K)}_{\Xi \Sigma^\ast} (q)
 +
 \frac{4}{3}
 \left( \frac{f_{\pi N \Delta}^\ast}{\sqrt{2} m_\pi} \right)^2
 \left( \frac{f_{\pi N \Delta}^\ast}{\sqrt{6} m_\pi} \right)^2
 V^{(C,\Kbar K)}_{\Xi^\ast \Sigma^\ast} (q)
\nonumber \\
&&
 +
 \left( \frac{3F-D}{2 \sqrt{3} f_\pi} \right)^2
 \left( \frac{3F+D}{2 \sqrt{3} f_\pi} \right)^2
 V^{(C,\Kbar K)}_{\Xi \Lambda} (q)
 +
 \frac{2}{3}
 \left( \frac{f_{\pi N \Delta}^\ast}{\sqrt{2} m_\pi} \right)^2
 \left( \frac{3F+D}{2 \sqrt{3} f_\pi} \right)^2
 V^{(C,\Kbar K)}_{\Xi^\ast \Lambda} (q)
\nonumber \\
\end{eqnarray}

In this case 
 the $K \Kbar$ crossed diagrams have more variety than
 for the $\pi \pi$ case, but the potential does not change 
 so much because the contribution of $V^{(C,K \Kbar)}$
 is quite small as explained before.

\section{The $\omega$ exchange contribution}
Here we take into account the $\omega$ exchange potential
 which is known as one of main source of short range repulsion
 of the baryon-baryon interaction.
The $\omega$ exchange potential for $\Lambda N$ 
 in momentum space is given by
\begin{eqnarray}
V_{\Lambda N}^\omega(q) 
 = 
  \frac{g_{\omega \Lambda \Lambda} g_{\omega NN}}
       {q^2+m_\omega^2}
  \left(
  \frac{\Lambda_\omega^2 - m_\omega^2}
       {\Lambda_\omega^2 + q^2}
  \right)^2
\label{EQ:POTOMEGA}
\end{eqnarray}
where we choose $g_{\omega NN}=13$ and
 $\Lambda_\omega = 1.4{\rm{GeV}}$ \cite{palomar}.
The ideal mixing for $\omega$ and $\phi$ leads 
 the relation,
 $g_{\omega \Lambda \Lambda} = \frac{2}{3} g_{\omega NN}$, 
 which is deduced from the quark contents of the hadrons.
For simplicity
 we assume the same form factor for both the $\omega NN$ and
 $\omega \Lambda \Lambda$ vertices.

\section{The isoscalar exchange 
         in the $\Lambda \Lambda$ interaction}
\begin{figure}
\begin{center}
 \includegraphics[scale=.85]{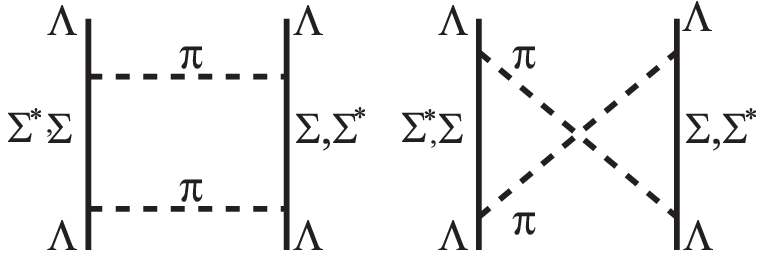} 
 \includegraphics[scale=.85]{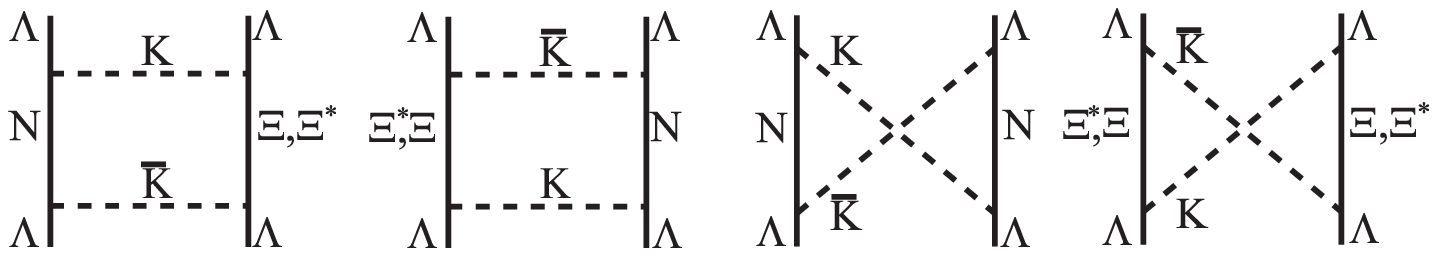}
\end{center}
\caption{Set of the uncorrelated two-meson exchange diagrams
         for the $\Lambda \Lambda$ interaction.}
\end{figure}
We easily extend this method to the $\Lambda \Lambda$ interaction.
For this
 we simply replace the triangle contribution of the nucleon by that
 of the $\Lambda$ in eq.~(\ref{EQ:Fullrep})
\begin{eqnarray}
 V^{Cor}_{\Lambda \Lambda} (q) 
  & = &
 \sum_{i,j}^{ \pi \pi, K \Kbar}
 N_{ij}
 \tilde{\Delta}^{i}_\Lambda ~
   t^{(I=0,L)}_{i \to j} ~
 \tilde{\Delta}^{j}_\Lambda 
\end{eqnarray}
where all contributions are already shown before.

The uncorrelated two-meson exchange potential is given by
\begin{eqnarray}
v^{(D,\pi \pi)}_{\Lambda \Lambda}
&=&
 3
 \left( \frac{D}{\sqrt{3} f_\pi} \right)^4
 V^{(D,\pi \pi)}_{\Sigma \Sigma} (q)
 + 
 3
 \left( \frac{f_{\pi N \Delta}^\ast}{\sqrt{2} m_\pi} \right)^2
 \left( \frac{D}{\sqrt{3} f_\pi} \right)^2
 V^{(D,\pi \pi)}_{\Sigma^\ast \Sigma} (q)
\nonumber \\
&&
 +
 3
 \left( \frac{D}{\sqrt{3} f_\pi} \right)^2
 \left( \frac{f_{\pi N \Delta}^\ast}{\sqrt{2} m_\pi} \right)^2
 V^{(D,\pi \pi)}_{\Sigma \Sigma^\ast} (q)
 + 
 3
 \left( \frac{f_{\pi N \Delta}^\ast}{\sqrt{2} m_\pi} \right)^4
 V^{(D,\pi \pi)}_{\Sigma^\ast \Sigma^\ast} (q)
\nonumber \\
v^{(C,\pi \pi)}_{\Lambda \Lambda}
&=&
 3
 \left( \frac{D}{\sqrt{3} f_\pi} \right)^4
 V^{(C,\pi \pi)}_{\Sigma \Sigma} (q)
 + 
 3
 \left( \frac{f_{\pi N \Delta}^\ast}{\sqrt{2} m_\pi} \right)^2
 \left( \frac{D}{\sqrt{3} f_\pi} \right)^2
 V^{(C,\pi \pi)}_{\Sigma^\ast \Sigma} (q)
\nonumber \\
&&
 +
 3
 \left( \frac{D}{\sqrt{3} f_\pi} \right)^2
 \left( \frac{f_{\pi N \Delta}^\ast}{\sqrt{2} m_\pi} \right)^2
 V^{(C,\pi \pi)}_{\Sigma \Sigma^\ast} (q)
 + 
 3
 \left( \frac{f_{\pi N \Delta}^\ast}{\sqrt{2} m_\pi} \right)^4
 V^{(C,\pi \pi)}_{\Sigma^\ast \Sigma^\ast} (q)
\end{eqnarray}

The two-kaon exchange is given by
\begin{eqnarray}
v^{(D,K \Kbar)}_{\Lambda \Lambda}
&=&
 2
 \left( \frac{3F-D}{2 \sqrt{3} f_\pi} \right)^2
 \left( \frac{D+3F}{2 \sqrt{3} f_\pi} \right)^2
 V^{(D,K \Kbar)}_{\Xi N} (q)
 +
 2
 \left( \frac{f_{\pi N \Delta}^\ast}{\sqrt{2} m_\pi} \right)^2
 \left( \frac{D+3F}{2 \sqrt{3} f_\pi} \right)^2
 V^{(D,K \Kbar)}_{\Xi\ast N} (q)
\nonumber \\
&&
 + 
 2
 \left( \frac{D+3F}{2 \sqrt{3} f_\pi} \right)^2
 \left( \frac{3F-D}{2 \sqrt{3} f_\pi} \right)^2
 V^{(D,\Kbar K)}_{N \Xi} (q)
 +
 2
 \left( \frac{D+3F}{2 \sqrt{3} f_\pi} \right)^2
 \left( \frac{f_{\pi N \Delta}^\ast}{\sqrt{2} m_\pi} \right)^2
 V^{(D,\Kbar K)}_{N \Xi\ast} (q)
\nonumber \\
v^{(C,K \Kbar)}_{\Lambda \Lambda}
&=&
 2
 \left( \frac{3F-D}{2 \sqrt{3} f_\pi} \right)^4
 V^{(C,K \Kbar)}_{\Xi \Xi} (q)
 + 
 2
 \left( \frac{f_{\pi N \Delta}^\ast}{\sqrt{2} m_\pi} \right)^2
 \left( \frac{3F-D}{2 \sqrt{3} f_\pi} \right)^2
 V^{(C,K \Kbar)}_{\Xi^\ast \Xi} (q)
\nonumber \\
&&
 + 
 2
 \left( \frac{3F-D}{2 \sqrt{3} f_\pi} \right)^2
 \left( \frac{f_{\pi N \Delta}^\ast}{\sqrt{2} m_\pi} \right)^2
 V^{(C,K \Kbar)}_{\Xi \Xi^\ast} (q)
 + 
 2
 \left( \frac{f_{\pi N \Delta}^\ast}{\sqrt{2} m_\pi} \right)^4
 V^{(C,K \Kbar)}_{\Xi^\ast \Xi^\ast} (q)
\nonumber \\
&&
 + 
 2
 \left( \frac{D+3F}{2 \sqrt{3} f_\pi} \right)^4
 V^{(C,\Kbar K)}_{N N} (q)
\end{eqnarray}

The $\omega$ exchange potential is also considered by
 substituting $g_{\omega NN}$ to
 $g_{\omega \Lambda \Lambda}$ in eq.~(\ref{EQ:POTOMEGA}).

\section{Scalar $\pi K$ exchange in the $\kappa$ channel}
In the interactions of the octet of mesons in the $\pi K$ channel
 one also finds a very broad resonance \cite{oop}, the kappa with
 $S=-1$, $I=1/2$, $J^P=0^+$, around $800~{\rm{MeV}}$, somewhat
 controversial \cite{Cherry:2000ut,Bugg:2006sz,
 Zhou:2006wm,Bugg:2005xx,Ablikim:2005ni}.
Its exchange is also accounted for in the recent model of
 \cite{haidenbauer}.

\begin{figure}
\begin{center}
 \includegraphics[scale=.85]{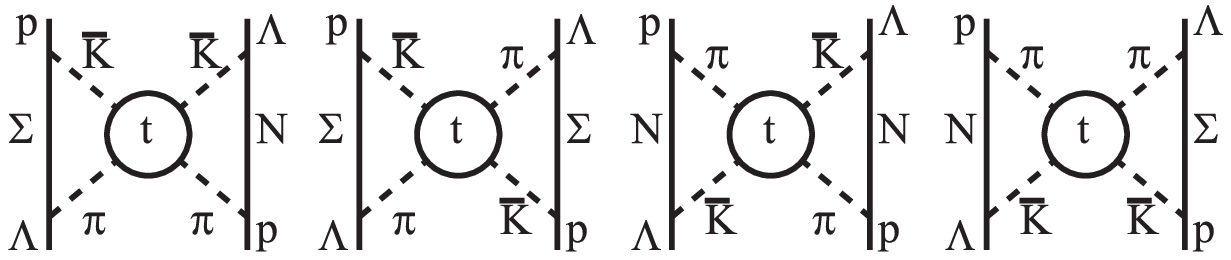} \\
 \includegraphics[scale=.85]{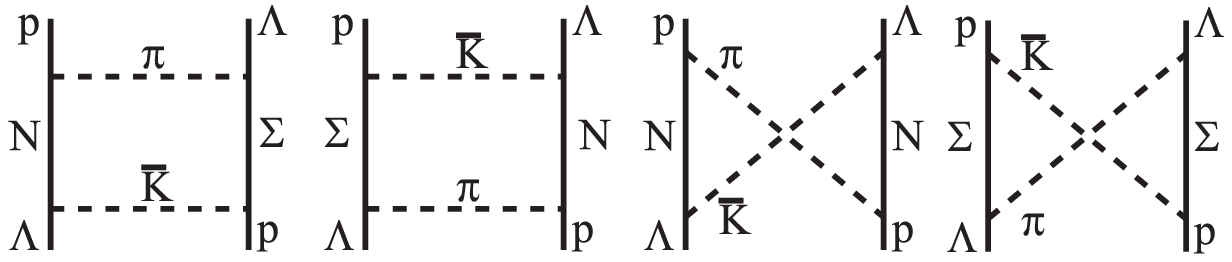}
\end{center}
\caption{Set of diagrams which contribute to the $\kappa$ channel.}
\label{Fig:5}
\end{figure}
Here we follow the same approach as in the former section and
 exchange $\pi K$ in the $I=1/2$, $l=0$ channel.
For this we consider the diagrams of Fig.~\ref{Fig:5}.
By following the same rules as before
 we find for the sum of all diagrams the compact expression
\begin{eqnarray}
 3~ t^{(I=1/2,L)}_{\pi \Kbar \to \pi \Kbar}
  \left\{
 - \left( \frac{D+3F}{2 \sqrt{3} f_\pi} \right)
   \left( \frac{D+F}{2 f_\pi} \right) V^{(\pi K)}_N (q)
 + \left( \frac{D}{\sqrt{3} f_\pi} \right)
   \left( \frac{D-F}{2 f_\pi} \right) V^{(\pi K)}_\Sigma (q)
  \right\}^2
\end{eqnarray}
which including the contribution of $\Sigma^\ast$ states gives
 (there are no $\Delta$ or $\Xi^\ast$ intermediate states now)
\begin{eqnarray}
 3~ t^{(I=1/2,L)}_{\pi \Kbar \to \pi \Kbar}
  \left\{
 - \left( \frac{D+3F}{2 \sqrt{3} f_\pi} \right)
   \left( \frac{D+F}{2 f_\pi} \right) V^{(\pi K)}_N (q)
 + \left( \frac{D}{\sqrt{3} f_\pi} \right)
   \left( \frac{D-F}{2 f_\pi} \right) V^{(\pi K)}_\Sigma (q)
  \right.
 \nonumber \\
  \left.
 - \frac{2}{3}
   \left( \frac{f_{\pi N \Delta}^\ast}{\sqrt{6} m_\pi} \right)
   \left( \frac{f_{\pi N \Delta}^\ast}{\sqrt{2} m_\pi} \right) 
    V^{(\pi K)}_{\Sigma^\ast} (q)
  \right\}^2
\end{eqnarray}

The $\pi \Kbar \to \pi \Kbar$ amplitude in the $I=1/2$ channel
 can be obtained from the appendix of \cite{oop} and we have 
\begin{eqnarray}
 t^{(I=1/2,L)}_{\pi \Kbar \to \pi \Kbar} 
  = 
 \frac{1}{4f_\pi^2}( 3u - s - 2 m_\pi^2 - 2 m_K^2 )
\end{eqnarray}
which after projection over $l=0$ gives \cite{Llanes-Estrada:2003us}
\begin{eqnarray}
 t^{(I=1/2,L)}_{\pi \Kbar \to \pi \Kbar}(l=0) 
  = 
 \frac{1}{4f_\pi^2}
 \left(
 - \frac{5}{2}s + m_\pi^2 + m_K^2 
 + \frac{3(m_K^2 - m_\pi^2)^2}{2s}
 \right).
\label{EQ:ApKk}
\end{eqnarray}
In order to avoid the singular behavior around $s=0$,
 we take the $SU(3)_f$ limit in the $\pi \Kbar$ amplitude.
Then
 we can obtain the modified $\pi \Kbar \to \pi \Kbar$ amplitude as
\begin{eqnarray}
 t^{(I=1/2,L)}_{\pi \Kbar \to \pi \Kbar}(l=0) 
  = 
 \frac{1}{4f_\pi^2}
 \left(
 - \frac{5}{2}s + 2 {m'}^2 
 \right)
\label{EQ:ApKkL}
\end{eqnarray}
where $m'$ is the average mass of pion and kaon.
This amplitude is also unitarized in the same way as before with
 only one channel.

Next 
 we calculate the uncorrelated $\pi K$ diagrams shown in
 Fig.~\ref{Fig:5} considering the decuplet excitation
 of the intermediate baryon.
These diagrams give
\begin{eqnarray}
v^{(D,\pi K)}_{\Lambda N(\kappa)}
& = &
 -3
 \left( \frac{D+F}{2 f_\pi} \right)
 \left( \frac{D+3F}{2 \sqrt{3} f_\pi} \right)
 \left( \frac{D}{\sqrt{3} f_\pi} \right)
 \left( \frac{D-F}{2 f_\pi} \right)
 V^{(D,\pi K)}_{N \Sigma} (q)
\nonumber \\
&&
 + \frac{1}{\sqrt{3}}
 \left( \frac{D+F}{2 f_\pi} \right)
 \left( \frac{D+3F}{2 \sqrt{3} f_\pi} \right)
 \left( \frac{f_{\pi N \Delta}^\ast}{m_\pi} \right)^2
 V^{(D,\pi K)}_{N \Sigma^\ast} (q)
\nonumber \\
v^{(D,K \pi)}_{\Lambda N(\kappa)}
& = &
 -3
 \left( \frac{D+F}{2 f_\pi} \right)
 \left( \frac{D}{\sqrt{3} f_\pi} \right)
 \left( \frac{D+3F}{2 \sqrt{3} f_\pi} \right)
 \left( \frac{D-F}{2 f_\pi} \right)
 V^{(D,K \pi)}_{\Sigma N} (q)
\nonumber \\
&&
 + \frac{1}{\sqrt{3}}
 \left( \frac{D+F}{2 f_\pi} \right)
 \left( \frac{D+3F}{2 \sqrt{3} f_\pi} \right)
 \left( \frac{f_{\pi N \Delta}^\ast}{m_\pi} \right)^2
 V^{(D,K \pi)}_{\Sigma^\ast N} (q)
\nonumber \\ 
v^{(C,\pi K)}_{\Lambda N(\kappa)}
& = &
 3
 \left( \frac{D+F}{2 f_\pi} \right)^2
 \left( \frac{D+3F}{2 \sqrt{3} f_\pi} \right)^2
 V^{(C,\pi K)}_{N N} (q)
\nonumber \\
v^{(C,K \pi)}_{\Lambda N(\kappa)}
& = &
 3
 \left( \frac{D}{\sqrt{3} f_\pi} \right)^2
 \left( \frac{D-F}{2 f_\pi} \right)^2
 V^{(C,K \pi)}_{\Sigma \Sigma} (q)
 -
 \frac{1}{\sqrt{3}}
 \left( \frac{D}{\sqrt{3} f_\pi} \right)
 \left( \frac{D-F}{2 f_\pi} \right)
 \left( \frac{f_{\pi N \Delta}^\ast}{m_\pi} \right)^2
 V^{(C,K \pi)}_{\Sigma^\ast \Sigma} (q)
\nonumber \\
&&
 -
 \frac{1}{\sqrt{3}}
 \left( \frac{D}{\sqrt{3} f_\pi} \right)
 \left( \frac{D-F}{2 f_\pi} \right)
 \left( \frac{f_{\pi N \Delta}^\ast}{m_\pi} \right)^2
 V^{(C,K \pi)}_{\Sigma \Sigma^\ast} (q)
 +
 \frac{1}{9}
 \left( \frac{f_{\pi N \Delta}^\ast}{m_\pi} \right)^4
 V^{(C,K \pi)}_{\Sigma^\ast \Sigma^\ast} (q)
\end{eqnarray}
where we follow this prescription for the meson pair of the
 superindex:
 the first meson corresponds to the upper one in the direct exchange 
 and to the upper one on the left baryon for the crossed 
 terms.

\section{Results}
\subsection{$\Lambda N$ Potential in momentum space}
\begin{figure}
 \begin{center}
  \begin{tabular}{cc}
   \includegraphics[scale=.47]{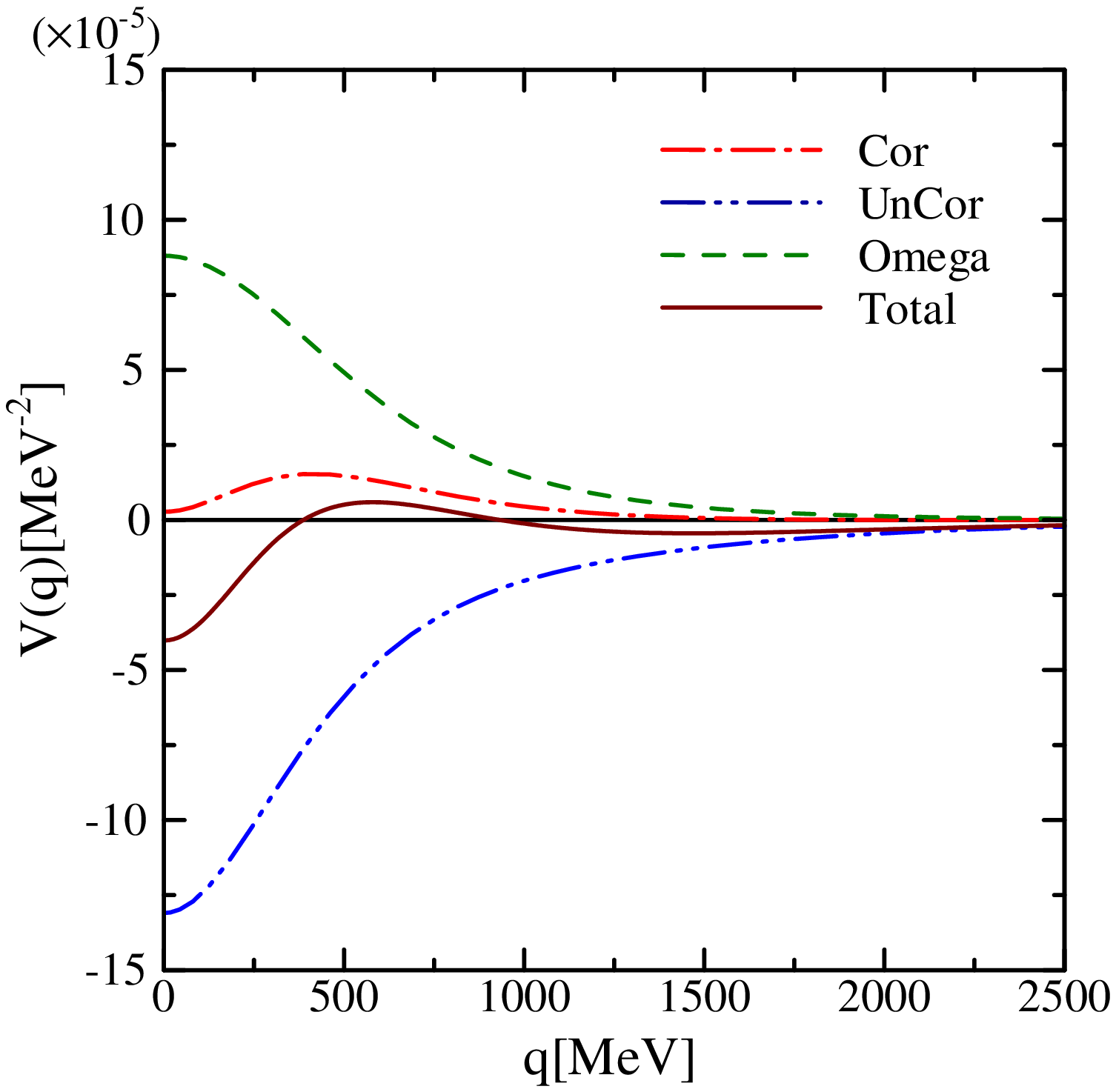} &
   \includegraphics[scale=.47]{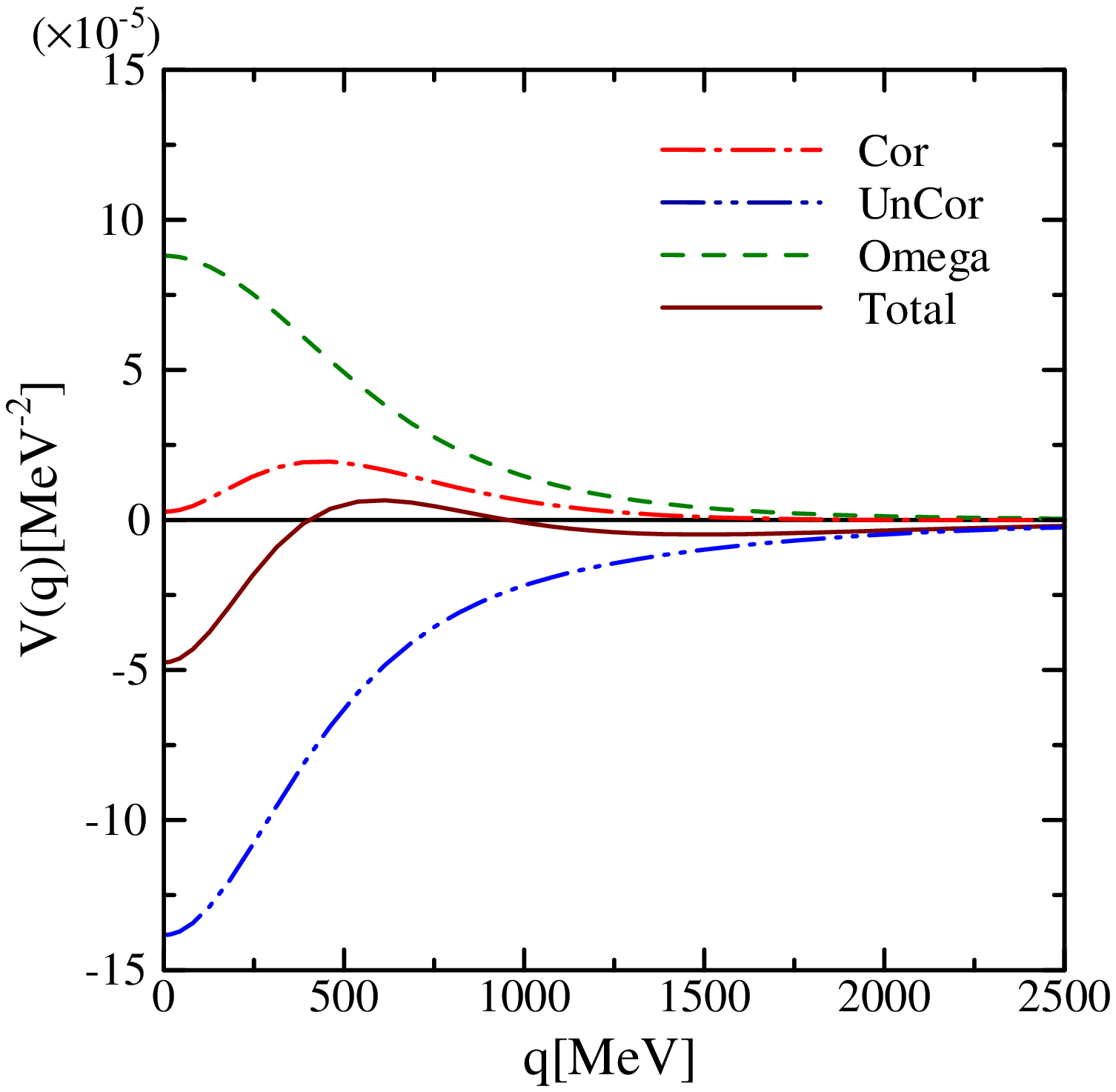} \\[-0.5em]
   (A) $\Lambda N$ potential with exchange of 
       pion pair &
   (B) $\Lambda N$ potential with exchange of 
       pion and kaon pairs \\[-1.5em]
  \end{tabular}
 \end{center}
\caption{The scalar-isoscalar $\Lambda N$ potential in momentum space.}
\label{FIG:PSPACEPOT}
\end{figure}
Fig.~\ref{FIG:PSPACEPOT} shows the $\Lambda N$ potential in
 momentum space.
The correlated two-meson contribution has a peak around
 $q=400~{\rm{MeV}}$.
A similar peak in position and magnitude was found for
 the $NN$ case in Refs.~\cite{osettoki,palomar,Kaskulov:2005kr}.
It is worth discussing this shape because it is impossible to
 parameterize it by a single meson exchange with usual form factors,
 like monopole or Gaussian.
This contribution could be decomposed in, at least, two parts,
 one a strong repulsive part and the other a weak attraction.
In any case, 
 this contribution is much smaller than the other ones,
 so that the main contribution comes from the uncorrelated 
 two-meson exchange and from $\omega$ exchange.
These two potentials have opposite sign, and the uncorrelated 
 two-meson potential is slightly stronger than the $\omega$
 exchange potential in the whole range of $q$.
Thus, the sum of these two potentials is always negative.


The total potential has positive strength around 
 $q=600~{\rm{MeV}}$ which is pushed up by the correlated
 two-meson potential.
The correlated two-meson potential plays a more important
 role in this region.

We can see the relevance of the two-kaon contribution to
 this potential by comparing the panels (A) and (B) 
 in Fig.~\ref{FIG:PSPACEPOT}.
The two-kaon contribution was studied since one kaon exchange
 is important in the $\Lambda N$ interaction.
However, the effect of the two-kaon contribution, suppressed
 due to its heavy mass, is very small and does not change much
 the pionic potential.

\subsection{$\Lambda N$ potential in the configuration space}
\begin{figure}
 \begin{center}
  \begin{tabular}{cc}
   \includegraphics[scale=.47]{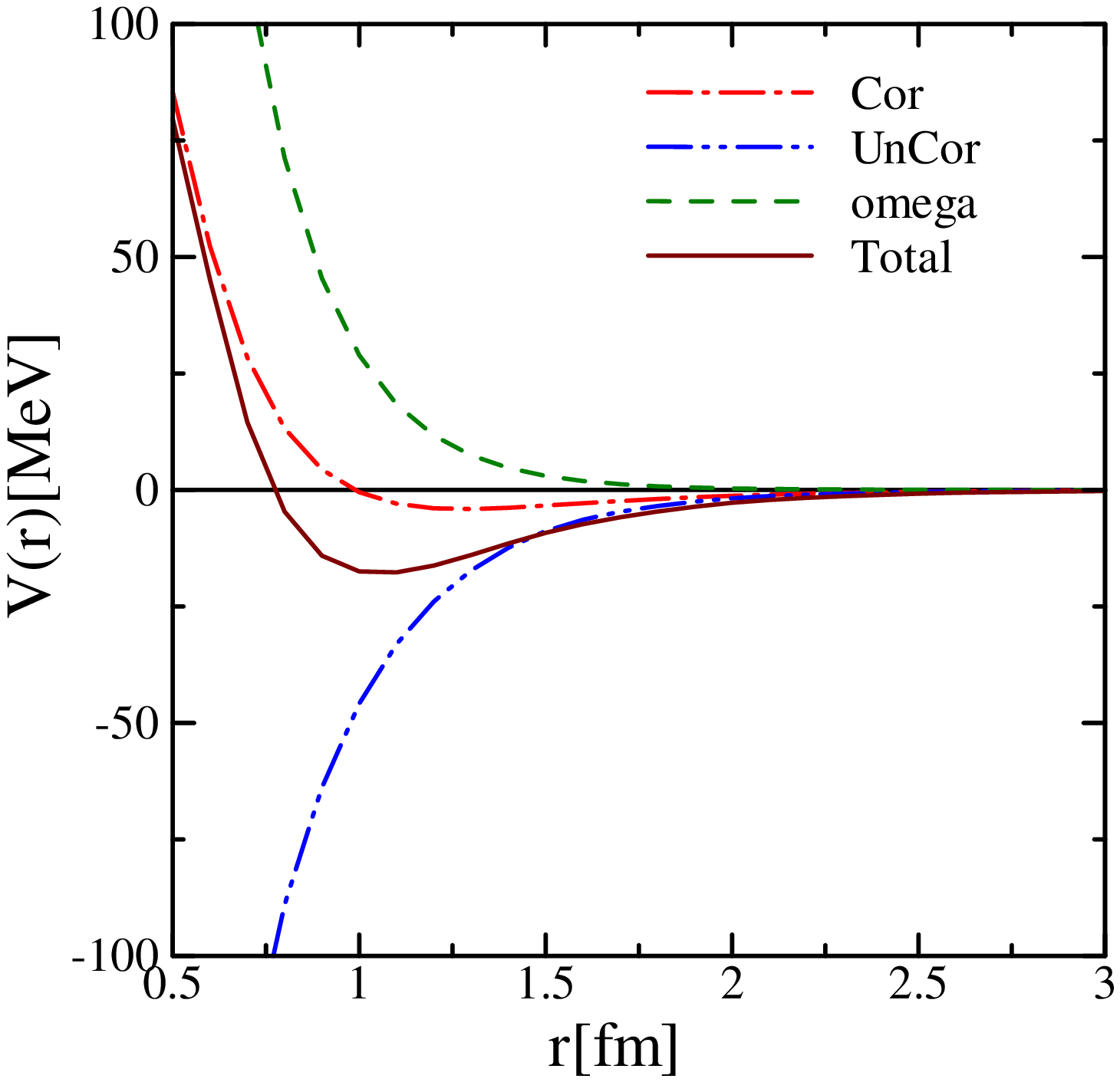} &
   \includegraphics[scale=.47]{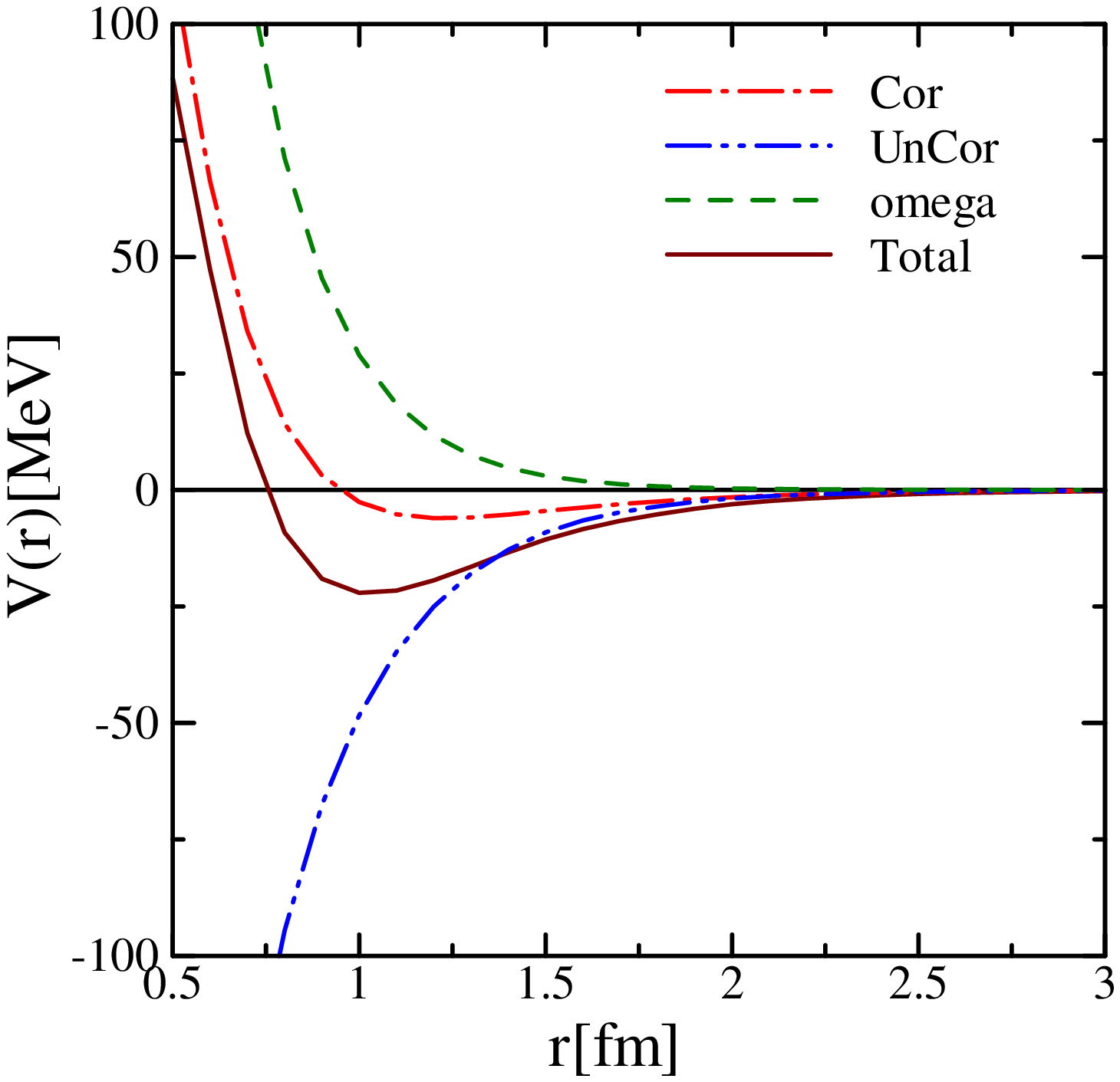} \\[-0.5em]
   (A) $\Lambda N$ potential with exchange of 
       pion pair &
   (B) $\Lambda N$ potential with exchange of 
       pion and kaon pairs \\[-1.5em]
  \end{tabular}
 \end{center}
\caption{The scalar-isoscalar $\Lambda N$ potential in 
         configuration space.}
\label{FIG:RSPACEPOT}
\end{figure}
The $\Lambda N$ potential in the configuration space is shown
 in Fig.~\ref{FIG:RSPACEPOT}.
The potential in configuration space is given by
\begin{eqnarray}
 V(r)
 = \frac{1}{2 \pi^2 r} \int_0^\infty q \sin(qr) V(q) dq .
\end{eqnarray}
In the left panel of this figure,
 we can see a similar correlated two-pion potential as for
 the $NN$ case, see Fig.~10 in \cite{osettoki}.
This should be expected since the definition of the potential is
 quite similar to the $NN$ case except for the masses of the 
 baryons and coupling constants. 
In fact 
 both the $NN$ and $\Lambda N$ potential generated by the 
 correlated two-pion exchange contribution
 pass through zero at $r \simeq 0.9~{\rm{fm}}$
 and have a minimum at $r \simeq 1.3~{\rm{fm}}$.
This potential is repulsive in the short range region
 and, on the other hand, it is attractive beyond
 $1~{\rm{fm}}$. 
As we have discussed in the previous subsection,
 the strength of the correlated two-meson potential is
 much smaller than the other contributions.

Fig.~\ref{FIG:RSPACEPOT} also shows that the uncorrelated two-meson
 generates a strong attraction and the $\omega$ produces
 a repulsion in the short range region.
The sum of these two potentials produces a relatively strong
 attraction around $1~{\rm{fm}}$ and leads to large cancellations
 in the short range region. 
We do not give the results below $0.5~{\rm{fm}}$ since there the
 overlap of the baryons and quark exchange mechanisms can 
 lead sizeable corrections.

Although the attraction in the total potential is mainly 
 generated by the uncorrelated two-meson potential, part of
 the repulsion is generated by the correlated two-meson potential.
This is interesting because the correlated two-meson potential
 is considered as a $\sigma$ meson exchange in other papers and,
 there, the interaction would
 be always attractive (see eq.~(3.19) of \cite{osettoki}).

Here, again, we can check the effect of two-kaon exchange potential
 by comparing the two panels in Fig.~\ref{FIG:RSPACEPOT}.
The two-kaon contribution slightly enhances the magnitude of
 both the correlated and uncorrelated two-meson potential,
 and it makes the total potential a little deeper than
 the pionic potential.



\subsection{The $\kappa$ exchange $\Lambda N$ potential}
\begin{figure}
 \begin{center}
  \begin{tabular}{cc}
   \includegraphics[scale=.47]{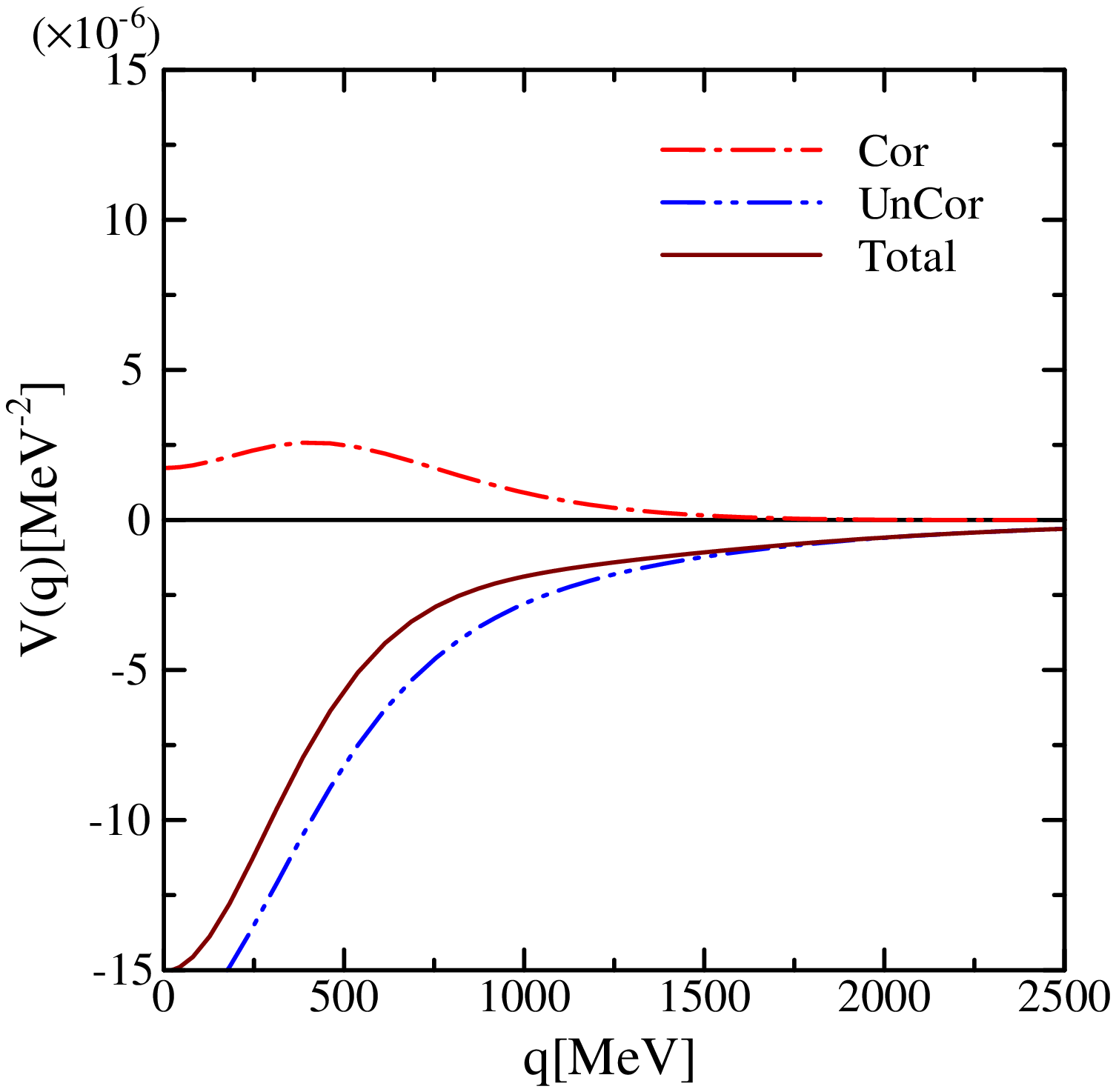} &
   \includegraphics[scale=.47]{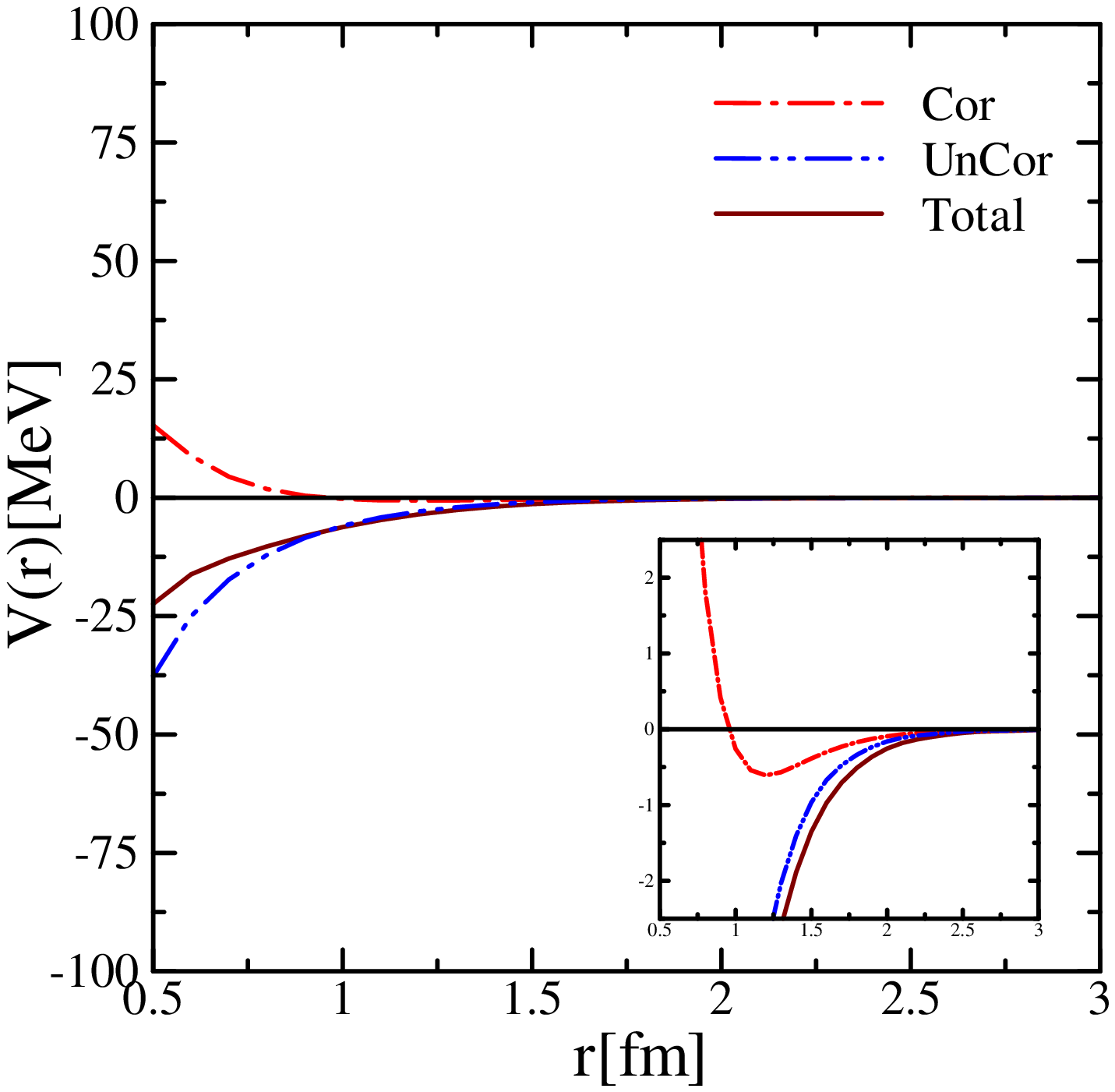} 
  \end{tabular}
 \end{center}
\caption{The $\Lambda N$ potential in $\kappa$ channel
         both for momentum space and configuration space.
         In the inset, we change the scale of the $V(r)$ axis
         to investigate the behavior of this potential.}
\label{FIG:KAPPA}
\end{figure}
Fig.~\ref{FIG:KAPPA} shows the $\kappa$ exchange contribution
 in the $\Lambda N$ potential.
The potential in momentum space is shown in the left panel of
 Fig.~\ref{FIG:KAPPA}. 
The correlated two-meson contribution has a similar shape
 as for the scalar-isoscalar channel, but its size
 is one order of magnitude smaller.

The uncorrelated two-meson exchange contribution generates
 an attraction which is relatively large compared to
 the correlated two-meson potential.

The right panel of Fig.~\ref{FIG:KAPPA} shows the potential
 in the configuration space.
The correlated two-meson contribution produces a moderate
 repulsion at the short range region and ends up at around
 $1.0~{\rm{fm}}$.
The uncorrelated two-meson potential produces a weak attraction 
 up to $2.0~{\rm{fm}}$.
Both potentials are of a quite short range compared to
 the isoscalar channel.
This reflects the heavy mass of the $\kappa$ meson.

The total $\kappa$ channel interaction is weakly
 attractive and of quite short range.
The strength of the potential is weaker than the isoscalar
 potential, almost $1/3$ at $1~{\rm{fm}}$.

\subsection{The $\Lambda \Lambda$ potential}
\begin{figure}
\begin{tabular}{cc}
\includegraphics[scale=.5]{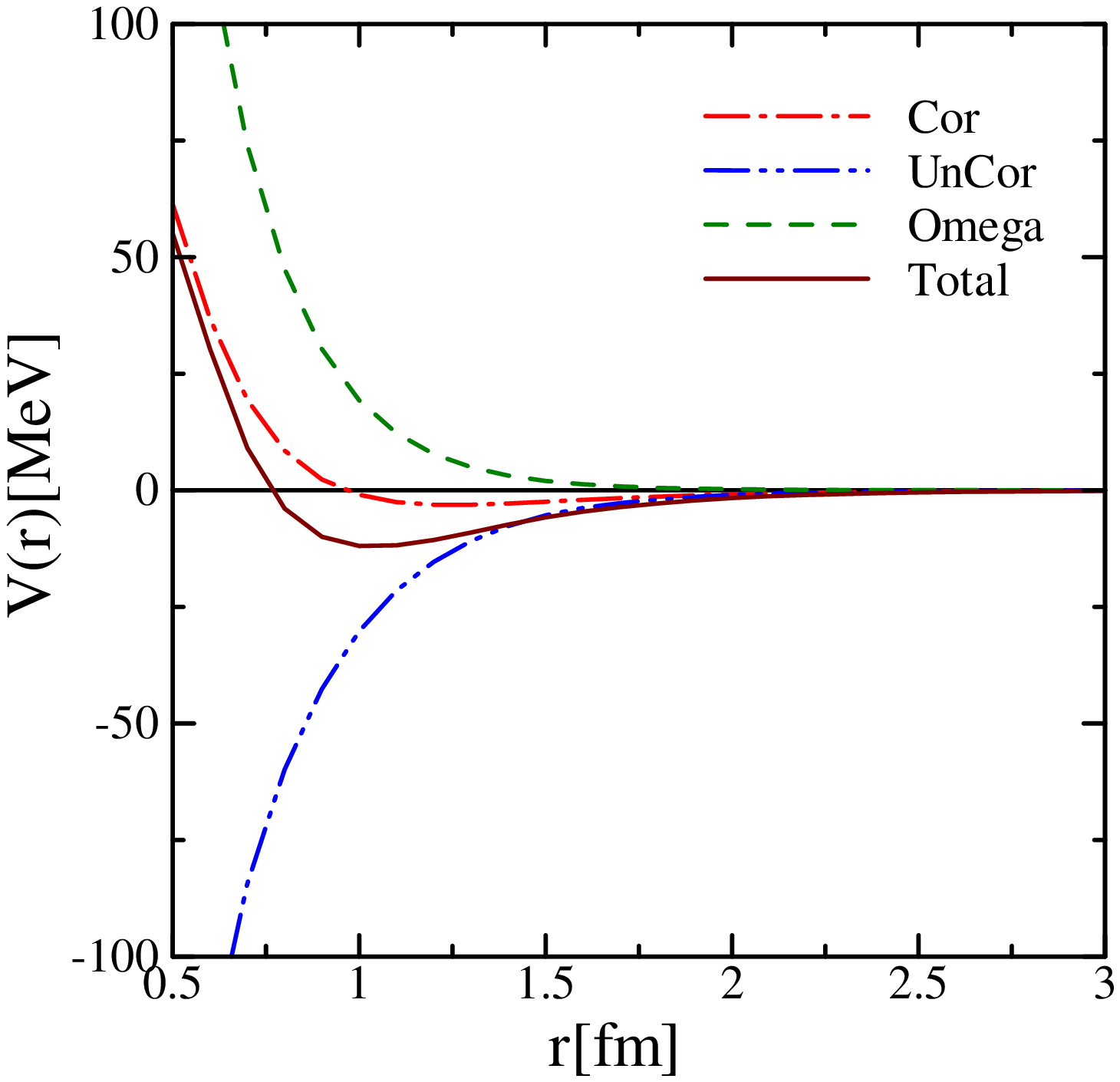}
\includegraphics[scale=.5]{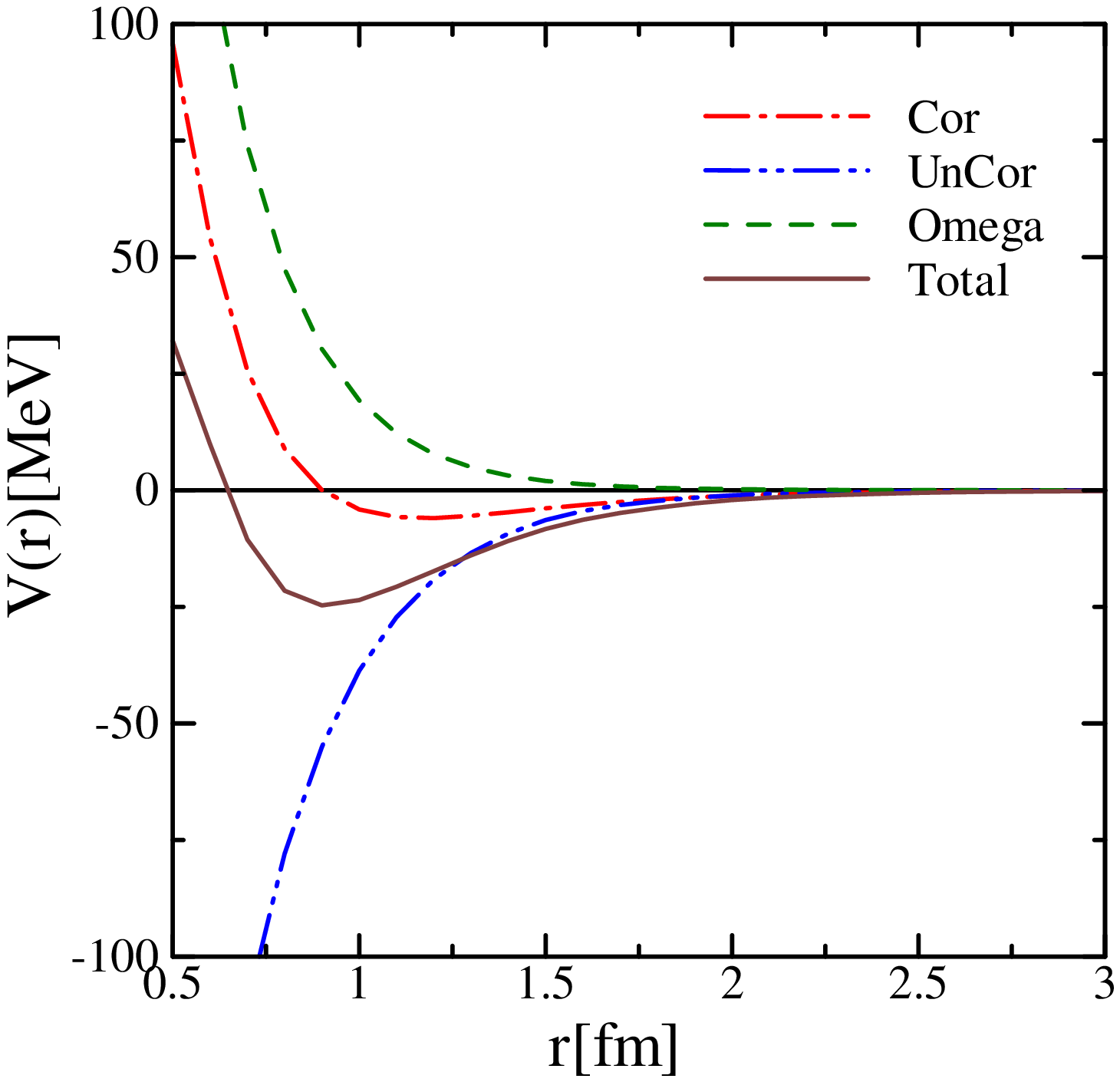}
\end{tabular}
\caption{The scalar-isoscalar $\Lambda \Lambda$ potential in 
         configuration space.}
\label{FIG:LLPOTC}
\end{figure}
Fig.~\ref{FIG:LLPOTC} shows the central part of
 the $\Lambda \Lambda$ potential.
Both the uncorrelated two-meson and the $\omega$ exchange
 contribution are weaker than for the $\Lambda N$ case,
 but they still have much larger magnitude than the correlated
 two-meson potential.  
These potentials drive the medium range attraction.
However,
 the $\omega$ exchange potential alone is not enough to produce 
 a repulsion in the shorter distances region.

We find again that the correlated two-meson potential 
 plays a dual role in making a repulsive potential in the short range
 region and a small attraction in the long range region.

These properties are seen both in the left and right panel cases.
The striking difference from the $\Lambda N$ interaction is 
 the large effect produced by the two-kaon exchange contribution.
Both the correlated and uncorrelated two-meson potential are
 largely enhanced by the two-kaon exchange diagrams especially
 in the shorter range region.
This effect leads to a large reduction of the short range repulsion. 
Consequently,
 the short range repulsion in the $\Lambda \Lambda$ potential is
 largely suppressed compared to the contribution of the two-pion
 exchange case and therefore to the $NN$ or $\Lambda N$
 interaction where the $K \Kbar$ exchange effect is much weaker.

\section{Conclusions}
We have evaluated the scalar channel potential between 
 the $\Lambda$ and nucleon.
We have considered the correlated and uncorrelated two-meson exchange
 contributions in this channel besides $\omega$ exchange.
The correlated two-meson exchange contribution was calculated
 by using a chiral unitary approach which reproduces very well
 the experimental meson-meson phase shift up to $1.2~{\rm{GeV}}$.

The uncorrelated two-meson exchange contribution produces
 a strong attraction which is similar to the $NN$ case.
The $\omega$ exchange contribution makes a short range repulsion,
 also similar to the $NN$ case, but its strength 
 is two-thirds of the $NN$ case due to the simple counting
 of non-strange quarks in the baryons.
These two contributions drive the attractive potential in the
 medium range region and almost cancel each other at shorter
 distances.

The correlated two-meson exchange contribution is relatively 
 smaller than the other two contributions.
This potential produces some attraction at medium range distances
 and some strong repulsion in the short range region.
This behavior is quite similar to the $NN$ case which is already
 calculated in \cite{osettoki}.
The striking effect is the repulsion in the short range region
 where the strong attraction generated by the uncorrelated
 two-meson potential is cancelled by the repulsion produced
 by the $\omega$ meson.
Thus the correlated two-meson potential plays an important role
 for both the medium range attraction and the short range 
 repulsion in the $\Lambda N$ and $\Lambda \Lambda$ interaction.

We have also checked the contribution of two-kaon exchange diagrams.
We have found that the two-kaon contribution is rather weak and 
 it slightly enhances the magnitude of the potential without changing
 its main behavior for the $\Lambda N$ potential. 
Therefore 
 it does not play an important role in the scalar $\Lambda N$ 
 potential.
On the other hand,
 it has a large contribution to the $\Lambda \Lambda$ potential,
 especially in the short distance region.
It largely enhances both the repulsion in the correlated two-meson
 exchange potential and the attraction in the uncorrelated one.
As a result,
 the total potential in the $\Lambda \Lambda$ interaction 
 becomes more attractive than for the $\Lambda N$ case and
 the short range repulsion is also reduced.

We have also investigated the $\pi K$ exchange in the $\kappa$ channel
 for the vase of the $\Lambda N$ interaction and it shows similar
 features to the scalar isoscalar potential for a shorter range
 and sizeably weaker strength.

Finally,
 we have found that the medium range attraction and short range
 repulsion is largely depend on the flavor of the baryons.
This flavor dependence of the central potential between baryons
 could be a clue to understand certain properties of nuclear
 structure and reactions.


\section*{Appendix}
\begin{table}
\caption{Particle assignment}
\begin{center}
\begin{tabular}{|c|c|}
\hline \hline
 Decuplet baryons &
 \begin{tabular}{cccc}
   $T^{111} = \Delta^{++}$ 
 & $T^{112} = \difrac{\Delta^{+}}{\sqrt{3}}$ 
 & $T^{122} = \difrac{\Delta^{0}}{\sqrt{3}}$
 & $T^{222} = \Delta^{-}$ \\[1.2em]
   $T^{113} = \difrac{\Sigma^{\ast +}}{\sqrt{3}}$ 
 & $T^{123} = \difrac{\Sigma^{\ast 0}}{\sqrt{6}}$ 
 & $T^{223} = \difrac{\Sigma^{\ast -}}{\sqrt{3}}$ & \\[1.2em]
   $T^{133} = \difrac{\Xi^{\ast 0}}{\sqrt{3}}$ 
 & $T^{123} = \difrac{\Xi^{\ast -}}{\sqrt{3}}$ & & \\[1.2em]
   $T^{333} = \Omega^-$ & & &
 \end{tabular} \\[1em]
\hline \hline
 Octet baryons &
 \begin{tabular}{ccc}
   $B^1_{\bar{1}} = \difrac{1}{\sqrt{6}} \Lambda 
                  + \difrac{1}{\sqrt{2}} \Sigma^{0}$ 
 & $B^1_{\bar{2}} = \Sigma^+$ 
 & $B^1_{\bar{3}} = p$ \\[1.2em]
   $B^2_{\bar{1}} = \Sigma^-$
 & $B^2_{\bar{2}} = \difrac{1}{\sqrt{6}} \Lambda 
                  - \difrac{1}{\sqrt{2}} \Sigma^{0}$
 & $B^2_{\bar{3}} = n$ \\[1.2em]
   $B^3_{\bar{1}} = \Xi^-$ 
 & $B^3_{\bar{2}} = \Xi^0$ 
 & $B^3_{\bar{3}} = -\sqrt{\difrac{2}{3}} \Lambda$ 
 \end{tabular} \\[1em]
\hline \hline
 Octet mesons &
 \begin{tabular}{ccc}
   $\Phi^1_{\bar{1}} = \difrac{1}{\sqrt{6}} \eta 
                     + \difrac{1}{\sqrt{2}} \pi^{0}$ 
 & $\Phi^1_{\bar{2}} = \pi^+$ 
 & $\Phi^1_{\bar{3}} = K^+$ \\[1.2em]
   $\Phi^2_{\bar{1}} = \pi^-$
 & $\Phi^2_{\bar{2}} = \difrac{1}{\sqrt{6}} \eta 
                  - \difrac{1}{\sqrt{2}} \pi^{0}$
 & $\Phi^2_{\bar{3}} = K^0$ \\[1.2em]
   $\Phi^3_{\bar{1}} = \Kbar^-$ 
 & $\Phi^3_{\bar{2}} = \Kbar^0$ 
 & $\Phi^3_{\bar{3}} = -\sqrt{\difrac{2}{3}} \eta$ 
 \end{tabular} \\[1em]
\hline \hline
\end{tabular}
\end{center}
\label{TAB:Asign}
\end{table}

After a non-relativistic reduction,
 the meson-baryon-baryon interaction with an emitted meson
 of momentum $\vec{q}$ is given by
\begin{eqnarray}
 -it^{Oct} 
  &=& 
  \left(
    \alpha_{MBB'} \frac{D+F}{2f_\pi} 
  + \beta_{MBB'}  \frac{D-F}{2f_\pi}
  \right)
  \vec{\sigma} \cdot \vec{q} 
\\
 -it^{Dec}
  &=& 
    \gamma_{MBB'} \frac{f_{\pi N \Delta}^\ast}{m_\pi}
  \vec{S}^\dagger \cdot \vec{q} 
\end{eqnarray}
where the $\sigma$ and $S$ are spin transition operators
 for the octet-octet and the octet-decuplet cases.
Here we define the $B' \to BM$ process.

\begin{table}
\begin{center}
\begin{tabular}{c|ccccc}
\hline \hline
$p \to$
 & $p \pi^0$
 & $n \pi^+$ 
 & $\Lambda K^+$
 & $\Sigma^0 K^+$
 & $\Sigma^+ K^0$
 \\
\hline
$\alpha_{MBp}$
 & $1$
 & $\sqrt{2}$
 & $-\difrac{2}{\sqrt{3}}$
 & $0$
 & $0$
 \\
\hline
$\beta_{MBp}$
 & $0$
 & $0$
 & $\difrac{1}{\sqrt{3}}$
 & $1$
 & $\sqrt{2}$
 \\
\hline
$p \to$
 & $\Delta^{++} \pi^-$
 & $\Delta^+ \pi^0$
 & $\Delta^0 \pi^+$
 & $\Sigma^{\ast +} K^0$
 & $\Sigma^{\ast 0} K^+$
 \\
\hline
$\gamma_{MBp}$
 & $1$
 & $-\sqrt{\difrac{2}{3}}$
 & $-\difrac{1}{\sqrt{3}}$
 & $\difrac{1}{\sqrt{3}}$ 
 & $-\difrac{1}{\sqrt{6}}$
 \\
\hline \hline
\end{tabular}

\begin{tabular}{c|ccccc}
\hline \hline
$n \to$
 & $p \pi^-$
 & $n \pi^0$ 
 & $\Lambda K^0$
 & $\Sigma^0 K^0$
 & $\Sigma^- K^+$
 \\
\hline
$\alpha_{MBn}$
 & $\sqrt{2}$
 & $-1$
 & $-\difrac{2}{\sqrt{3}}$
 & $0$
 & $0$
 \\
\hline
$\beta_{MBn}$
 & $0$
 & $0$
 & $\difrac{1}{\sqrt{3}}$
 & $-1$
 & $\sqrt{2}$
 \\
\hline
$n \to$
 & $\Delta^+ \pi^-$
 & $\Delta^0 \pi^0$
 & $\Delta^- \pi^+$
 & $\Sigma^{\ast 0} K^0$
 & $\Sigma^{\ast -} K^+$
 \\
\hline
$\gamma_{MBn}$
 & $\difrac{1}{\sqrt{3}}$
 & $-\sqrt{\difrac{2}{3}}$
 & $-1$
 & $\difrac{1}{\sqrt{6}}$ 
 & $-\difrac{1}{\sqrt{3}}$
 \\
\hline \hline
\end{tabular}

\begin{tabular}{c|ccccccc}
\hline \hline
$\Lambda \to$
 & $\Sigma^+ \pi^-$
 & $\Sigma^0 \pi^0$
 & $\Sigma^- \pi^+$
 & $p K^-$
 & $n \Kbar^0$
 & $\Xi^0 K^0$
 & $\Xi^- K^+$
 \\
\hline
$\alpha_{MB \Lambda}$ 
 & $\difrac{1}{\sqrt{3}}$
 & $\difrac{1}{\sqrt{3}}$
 & $\difrac{1}{\sqrt{3}}$
 & $-\difrac{2}{\sqrt{3}}$
 & $-\difrac{2}{\sqrt{3}}$
 & $\difrac{1}{\sqrt{3}}$
 & $\difrac{1}{\sqrt{3}}$ 
 \\
\hline
$\beta_{MB \Lambda}$ 
 & $\difrac{1}{\sqrt{3}}$
 & $\difrac{1}{\sqrt{3}}$
 & $\difrac{1}{\sqrt{3}}$
 & $\difrac{1}{\sqrt{3}}$
 & $\difrac{1}{\sqrt{3}}$
 & $-\difrac{2}{\sqrt{3}}$
 & $-\difrac{2}{\sqrt{3}}$ 
 \\
\hline
$\Lambda \to$ 
 & $\Sigma^{\ast +} \pi^-$
 & $\Sigma^{\ast 0} \pi^0$ 
 & $\Sigma^{\ast -} \pi^+$ 
 & $\Xi^{\ast 0} K^0$
 & $\Xi^{\ast -} K^+$
 & 
 & 
 \\
\hline
$\gamma_{MB \Lambda}$ 
 & $-\difrac{1}{\sqrt{2}}$ 
 & $\difrac{1}{\sqrt{2}}$
 & $\difrac{1}{\sqrt{2}}$ 
 & $-\difrac{1}{\sqrt{2}}$
 & $\difrac{1}{\sqrt{2}}$ 
 &
 &
 \\
\hline \hline
\end{tabular}
\end{center}
\end{table}

\section*{Acknowledgments}
One of us, K.S., wishes to acknowledge support from
 the Ministerio de Educaci\'on y Ciencia and program of
 estancias de j\'ovenes doctores y tecn\'ologos extranjeros
 en Espa\~na. 
This work is partly supported by DGICYT contract number BFM2003-00856,
and the E.U. EURIDICE network contract no. HPRN-CT-2002-00311.
This research is part of the EU Integrated Infrastructure Initiative
Hadron Physics Project under contract number RII3-CT-2004-506078.

\end{document}